\documentclass[12pt]{article}

\usepackage{amsmath,amssymb,epsfig,axodraw}
\begin{document}

\hfill{SNUTP 04-021}
\begin{center}
{\Large{\bf  Orbifold Compactification and Related
Phenomenology\footnote{Talk given at SI2004, Fuji-Yoshida, Japan,
August 12-19, 2004 (to be published in the Proceedings).}}}
\end{center}
\vskip .5 truecm \centerline{\bf Kyuwan Hwang and Jihn E. Kim}
\vskip .4 truecm \centerline {\it School of Physics and Center for
Theoretical Physics} \centerline {\it Seoul National University,
Seoul 151-747, Korea} \vskip 0.5 truecm

\makeatletter \@addtoreset{equation}{section}
\def\theequation{\thesection.\arabic{equation}}
\makeatother

\begin{abstract}
In this lecture, starting with GUT we go through the field
theoretic and stringy orbifold compactification of extra
dimensions toward their application to low energy particle
physics.
\end{abstract}

\def\sw0{$\sin^2\theta_W^0$}
\def\wa{{$\sin^2\theta_W$}}

\def\Tr{{\rm Tr}}
\def\L{{\rm L}}
\def\R{{\rm R}}
\def\Z{{\bf Z}}

\def\Nr{{\rm N}}
\def\N{${\cal N}$}

\def\p{\partial}
\def\one{\bf 1}
\def\two{\bf 2}
\def\five{\bf 5}
\def\ten{\bf 10}
\def\tenb{\overline{\bf 10}}
\def\fiveb{\overline{\bf 5}}
\def\threeb{{\bf\overline{3}}}
\def\three{{\bf 3}}
\def\ts{{\bf 27}}
\def\tsb{$\overline{\bf 27}$}
\def\fb{{\overline{F}\,}}
\def\hb{{\overline{h}}}
\def\Hb{{\overline{H}\,}}

\def\hf{\textstyle{\frac12~}}
\def\hff{\textstyle{\frac13~}}
\def\hfg{\textstyle{\frac23~}}

\def\E88{{E$_8\times$E$_{8}^\prime$}}
\def\Esix{{E$_6$}}
\def\sw0{{$\sin^2\theta_W^0$}}
\def\slminus{\hskip -0.2cm\slash}

\section{Introduction}

In the last three decades, we observed that the standard model(SM)
is very successful phenomenologically. But the family problem is
not understood in the SM. The family problem addresses: (i) ``Why
are there 3 families?" and (ii) ``How does the Yukawa texture
arise?" In this talk, let us concentrate on Question (i) only. It
can be stated in other words as, ``Is 3 a very fundamental number
in the universe?"

A very good family unification model seems to be an SO(4$n$+2)
grand unification theory(GUT) with one spinor representation, but
a naive breaking of the SO(4$n$+2) to SO(10) does not lead to
chiral fermions. Shortly we will see how this dilemma can be
avoided.

The ``$V-A$" charged current(CC) weak interaction is described by
the left-handed weak doublets \cite{Weinberg67},
\begin{equation}\label{efamily}
l_L\equiv \left(
\begin{array}{c}
\nu_e\\ e
\end{array}
\right)_L,\ \ e_R\ ,\ \ q_L\equiv\left(
\begin{array}{c}
u^\alpha\\ d^\alpha
\end{array}
\right)_L,\ \ u_R^\alpha,\ \ d_R^\alpha
\end{equation}
which can be considered as a unification of the weak and
electromagnetic forces since $u_L$ and $d_L$ are put in the same
representation. Generalizing this, we adopt the unification
principle in this talk,

{\bf Theme of unification}: Put matter representations in one
representation in a bigger group.

GUT started by observing that all fermions of the SM can be put
into one chirality, e.g. to left-handed fields, which is Georgi
and Glashow's great contribution \cite{GG73}. ``Can we put 15
chiral fields of (\ref{efamily}) in a single representation?" We
know that it is possible by adding a singlet. This gives the
SO(10) GUT \cite{Georgiso10}. But, if we stick to just 15 chiral
fields, then $\ten+\fiveb$ is the simplest choice which is the
SU(5) GUT. The theme of unification prefers the SO(10) group.

There are at least 45(+3) chiral fields. Then, one may consider a
big group such as SU(45) or some SU(N) giving 45 chiral fields.
Indeed, this line of unification was considered in terms of
SU(11), SU(8), SU(9), etc. In these big GUTs, the survival
hypothesis \cite{Georgi79} is the guiding principle. But, the
anomaly problem exists in SU(N) groups. Except the SU(N) series,
there is no anomaly problem. Thus, SO(4$n$+2) with one complex
spinor, which has the dimension $2^{2n}$, seems to be a good
candidate for family unification. Its branching rule to
SO(4($n$-1)+2) is ${\bf 2^{2n}}=2\cdot[\bf 2^{2(n-1)}+
2^{2(n-1)*}].$ Thus, by applying the survival hypothesis a naive
breaking does  not lead to chiral fermions at low energy.

A method to obtain chiral fermions from the spinor of SO(4$n$+2)is
twisting\footnote{Twisting will be used later on torus.} the group
space \cite{Kim7}. Consider SO(14) with a spinor {\bf 64}. The
spinor branches to $\bf 35+\overline{21}+7+1$ under
SO(14)$\to$SU(7). By twisting, we mean that the electromagnetic
charge operator on SU(7) fundamental representation {\bf 7} is
taken as diag.$(-\frac13,-\frac13,-\frac13,1,0,1,-1)$ instead of
diag.$(-\frac13,-\frac13,-\frac13,1,0,0,0)$. The last two entries
signals the twisting. Without twisting, there appear two sets of
left-handed doublets and two sets of right-handed doublets, and
there results no family. But with the above twisting, the
electromagnetic charge of one set of the right-handed doublets are
shifted by $+1$ unit and the charges of the other set are shifted
by $-1$ unit, i.e. the doublets are
\begin{eqnarray}
&& \left(
\begin{array}{c}
\nu_e\\ e
\end{array}
\right)_L,\ \ \left(
\begin{array}{c}
u\\ d \end{array} \right)_L,\ \ \left(
\begin{array}{c}
\nu_\mu\\ \mu
\end{array}
\right)_L,\ \ \left(
\begin{array}{c}
c\\ s
\end{array}
\right)_L,\ \ \nonumber\\
\label{skewdrep}\\
&& \left(
\begin{array}{c}
\tau^+\\ \bar\nu_\tau
\end{array}
\right)_R,\ \ \left(
\begin{array}{c}
q_{5/3}\\ t \end{array} \right)_R,\ \ \left(
\begin{array}{c}
E^-\\ E^{--}
\end{array}
\right)_R,\ \ \left(
\begin{array}{c}
b\\ q_{-4/3}
\end{array}
\right)_R \nonumber
\end{eqnarray}
where the 32 charge-conjugated SU(2)-singlet fields are not shown.
It describes three lepton doublets and two quark doublets
correctly. But the third family quark doublet is not the one
observed. In addition, there appear strangely charged particles,
which will be encountered frequently in standard-like models in
orbifold compactification.

After the advent of string models in 1984, E$_8$ group has been
known to be big enough to house all the known fermions. If we
pursue on the SO(4$n$+2) route, we do not have a rationale for ``
why one spinor?". This is the representation problem in SO(4$n$+2)
GUTs. With the \E88\ and SO(32) heterotic string models with  the
adjoint representation only, there is no such representation
problem in string models.

In this talk, we start with field theoretic orbifold and introduce
the heterotic string and string orbifolds. Then, in the last
chapter we present a trinification GUT from orbifold
compactification of the \E88\ hetrotic string.

\section{Field theoretic orbifolds}

 Before we discuss string orbifolds, let us introduce field
 theoretic orbifolds which is easier to understand.
Here also, the objective is to obtain chiral fermions. Note that
in even dimensions we have to worry about anomalies in field
theoretic orbifolds.

\begin{figure}[h]
\begin{center}
\begin{picture}(150,150)(0,-20)
\BCirc(-40,50){40} \GCirc(-40,90){3}{0} \GCirc(-40,10){3}{0}
\DashLine(-40,-5)(-40,105){4} \LongArrow(-45,110)(-35,110)
\LongArrow(-35,110)(-45,110)

\DashLine(-75,65)(-5,65){4} \LongArrow(-8,65)(-4,65)
\LongArrow(-72,65)(-76,65)

\DashLine(-65,35)(-15,35){4} \LongArrow(-8,35)(-4,35)
\LongArrow(-72,35)(-76,35) \Text(-40,120)[]{$\Z_2$}
\Text(-33,97)[]{$0$} \Text(-27,3)[]{$\pi R$} \SetWidth{1.5}
\CArc(-40,50)(40,270,90)

\Text(-40,-25)[c]{(a)} \Text(190,-25)[c]{(b)}

\SetWidth{0.5} \DashLine(190,-5)(190,105){4} \BCirc(190,50){40}
\LongArrow(185,110)(195,110) \LongArrow(195,110)(185,110)
\DashLine(135,50)(245,50){4} \LongArrow(250,45)(250,55)
\DashLine(190,-5)(190,105){4} \LongArrow(250,55)(250,45)
\GCirc(190,90){3}{0} \GCirc(230,50){3}{0} \Text(190,120)[]{$\Z_2$}
\Text(265,50)[]{$\Z_2'$} \Text(197,97)[]{$0$} \Text(203,3)[]{$\pi
R$} \Text(239,40)[]{${\pi R \over 2}$} \SetWidth{1.5}
\CArc(190,50)(40,0,90)
\end{picture}
\end{center}
\caption{ (a) $S^1/\Z_2$ orbifold. (b) $S^1/(\Z_2\times \Z_2')$
orbifold. Fixed points at $y=0,{\pi R\over 2}$ are denoted by
bullets. The thick arc is the fundamental
region.\index{fundamental region}\label{s1z2z2'}}\label{s1z2}
\end{figure}
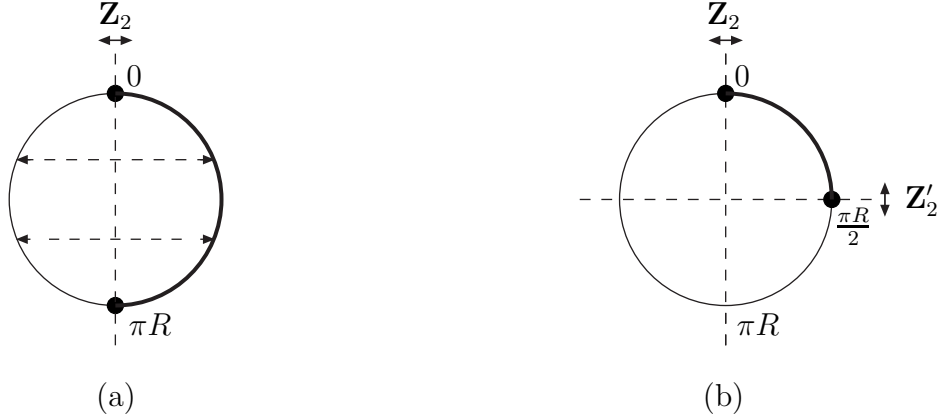

Consider a 1D torus, i.e. a circle. On this torus, the orbifold is
defined as the line segment $I=S^1/\Z_2$, which is the simplest
orbifold shown in Fig. \ref{s1z2},
$$\Z_2:\ \
y \longrightarrow -y
$$ where $S^1$ is coordinatized by $-\pi R < y
\le \pi R$.

This action has two {\it fixed points} $y=0$ and $y=\pi R$. These
fixed points stay there when the $\Z_2$ action is performed. The
fundamental region is defined as $[0,\pi R]$ which is shown as a
thick arc. In this one dimensional example, the two fixed points
are clearly singular even in the context of topology as well as in
differential geometry. Many examples in higher dimensions leave
the final spaces by this {\it orbifolding} with the same topology
as the original ones, but with fixed points which are singular in
differential geometry.

Let us start with the simplest five dimensional field theory model
\cite{Kawamura}. The orbifold along the fifth dimension is
$S^1/\Z_2$. Let the coordinate of the five dimensional(5D)
spacetime be $(x^\mu, y)$, where $x^\mu$ is for 4D and $y$ is the
fifth dimensional coordinate. We call the five dimensional space
as {\it bulk} while the surfaces on the end points of the line
segment $y=0$ and $y=\pi R$ as {\it branes}. The action of an
abelian gauge theory on the $M^4\times S^1/\Z_2$ is
\begin{equation}
\textstyle S = \int d^4x dy \ \left\{-{1\over 4}F^{MN} F_{MN}
  + \bar{\Psi} \Gamma^M (i\partial_M - gQ A_M)  \Psi\right\} \,,
  \end{equation}

Before orbifolding, i.e. imposing the $\Z_2$ boundary conditions,
the Kaluza-Klein(KK) mode expansion is, when compactified on
$S^1$,
\begin{equation}
\Psi(x,y) = \sum_{n=0}^{\infty} \Psi^{(n)}(x) \cos\left({ny\over
R}\right) + \sum_{n=1}^{\infty} \Psi^{c(n)}(x) \sin\left({ny\over
R}\right)\,.
\end{equation}
Using the chiral representation of gamma matrices where $\gamma^5$
is given by
\begin{equation}
\gamma^5 = \left(\begin{array}{cc}1 & 0\\ 0 & -1\end{array}\right)
\end{equation}
the consistency condition under the $\Z_2$ reflection $y
\longrightarrow -y$ requires $\psi$ and $\psi^c$ having different
eigenvalues\footnote{From the 5D Dirac equation, $\Z_2=\tilde \Z_2
\gamma^5$ where $\tilde \Z_2$ is just the operation $y\rightarrow
-y$.} under $\Z_2$:
\begin{equation} \label{z2bc}
\Z_2\ :\  \psi(x,y) \longrightarrow \eta\psi(x,-y), \quad
\psi^c(x,y) \longrightarrow -\eta\psi^c(x,-y),
\end{equation}
where an overall factor $\eta$ can be 1 or $-1$ and the
eigenvalues are $\eta$ and $-\eta$. For the time being, let us
choose $\eta= 1$. Then $\psi$ is even under $\Z_2$ while $\psi^c$
is odd under $\Z_2$, allowing only a part of the KK decomposition,
\begin{equation}
\psi(x,y) = \sum_{n=0}^{\infty} \psi^{(n)}(x)
    \cos\left( {ny\over R} \right),\quad
\psi^c(x,y) = \sum_{n=1}^{\infty} \psi^{c(n)}(x)
    \sin\left( {ny \over R}\right).\label{KKdecomp}
\end{equation}

Note that this property of obtaining an effective 4D chiral theory
is a direct consequence of $\Z_2$ symmetry, realized in the wave
function through consistently imposing boundary conditions. This
shows the easiest way of compactification toward an effective 4D
chiral gauge theory.

\vskip 0.3cm \noindent{\bf Gauge symmetry breaking}: The orbifold
$S^1/\Z_2$ we discussed in the previous section can be further
orbifolded by $\Z_2'( y \longrightarrow \pi R - y)$. Defining
$y^\prime=\frac{\pi R}{2}-y$, the $\Z_2^\prime$ action is, viz.
Fig. \ref{s1z2}(b),
\begin{equation}
\Z_2^\prime\ :\ \ y^\prime \longrightarrow -y^\prime.
\end{equation}
Since the $\Z_2^\prime$ action does not transform any point of the
thick arc in Fig. \ref{s1z2}(b) to the other half, $\Z_2^\prime$
is said to commute with $\Z_2$. Thus, the moding discrete group is
$\Z_2\times \Z_2^\prime$. In fact, it is the only way to introduce
another commuting discrete action on top of the existing $\Z_2$
when the number of extra-dumension is one. The $\Z_2'$ action is
the reflection with respect to the line connecting the two points
$y = \pi R/2$ and $y=-\pi R/2$.

In general, any bulk field $\phi(x^\mu,y)$ can have components
with the intrinsic $\Z_2\times \Z_2'$ parities of $(++),(+-),(-+)$
and $(--)$. Denoting those fields as
$\phi_{++},\phi_{+-},\phi_{-+}$ and $\phi_{--}$, they have mode
expansions in terms of sines and cosines with the following
$\Z_2\times \Z_2'$ parities

\begin{eqnarray}
\phi_{++}(x^\mu, y) &=& \sum_{n=0}^{\infty}
   \phi_{++}^{(2n)}(x^\mu) \cos {2ny\over R},\label{KK_ZZ1}  \\
\phi_{+-}(x^\mu, y) &=& \sum_{n=0}^{\infty}
   \phi_{+-}^{(2n+1)}(x^\mu) \cos {(2n+1)y\over R}, \label{KK_ZZ2} \\
\phi_{-+}(x^\mu, y) &=& \sum_{n=0}^{\infty}
   \phi_{-+}^{(2n+1)}(x^\mu) \sin {(2n+1)y\over R},  \label{KK_ZZ3}\\
\phi_{--}(x^\mu, y) &=& \sum_{n=0}^{\infty}
   \phi_{--}^{(2n+2)}(x^\mu) \sin {(2n+2)y\over R}\,.\label{KK_ZZ4}
\end{eqnarray}
Note that only $\phi_{++}$ can have a zero mode, allowing a
massless 4D field in the effective low energy theory.

For $\Z_2'$ being a reflection in the fifth coordinate, the
fermionic bulk fields should satisfy the second boundary
conditions similar to Eq.~(\ref{z2bc}), with different overall
factor $\eta^\prime$ :
\begin{equation} \label{z2pbc}
\Z_2'\ : \ \psi(x,y^\prime) \longrightarrow
\eta'\psi(x,-y^\prime), \quad \psi^c(x,y^\prime) \longrightarrow
-\eta'\psi^c(x,-y^\prime)\,.
\end{equation}
In order to obtain a chiral theory, one $\Z_2$ was enough as
discussed above. Now, we can use the additional $\Z_2'$ in
breaking another continuous symmetry of the system, especially the
nonabelian gauge symmetry \cite{Hebecker}. As the simplest
example, let us consider a toy model with an SU(3) gauge symmetry.
The action is given by
\begin{eqnarray}
S &=& \int d^4x dy \ \Big\{\textstyle -{1\over 4}{\rm
Tr}\left(F^{MN} F_{MN} \right)
  + \bar{\Psi} \Gamma^M (i\partial_M - g A_M)  \Psi  \nonumber \\
 &&\textstyle \quad\quad\quad\quad
   +\ \delta(y){\cal L}(x)
   + \delta\left( y - {\pi R\over 2} \right) {\cal L'}(x)\Big\}\,
\end{eqnarray}
where $A_M$ and $F_{MN}$ are Lie-algebra valued. $\cal L$ and
$\cal L'$ are the brane interactions which are localized at $y=0$
and $y=\pi R/2$ branes, respectively.

In order to break the gauge symmetry SU(3), we choose the boundary
condition under $\Z_2'$ as
\begin{equation} \label{gsbc}
A_M(x,y^\prime) =  \Lambda_M^N P A_N(x,-y^\prime) P^{-1}, \quad
\Psi(x,y^\prime) =  \gamma_5 P \Psi(x,-y^\prime),
\end{equation}
where $\Lambda = {\rm diag}(1,1,1,1,-1)$ and $P$ is given by
\begin{equation}
P = \left(\begin{array}{ccc}
 -1&  0&  0\\ 0&  -1&  0\\ 0& 0& 1\end{array}
\right) .
\end{equation}
Under this specific choice of $\Z_2'$, $A^{1,2,3}_\mu$ and
$A^8_\mu$ are even while $A^{4,5,6,7}_\mu$ are odd. Thus the 4D
gauge symmetry is broken down to SU(2)$\times $U(1) since the
SU(3)/(SU(2)$\times $U(1)) gauge fields do not have zero modes.
Breaking of the gauge symmetry can be easily noticed from
$[P,A_M]\ne 0$. Generally speaking, when the boundary conditions
Eq.(\ref{gsbc}) for an arbitrary $P$ is imposed, only those
subgroups commuting with $P$ survive.

\vskip 0.3cm \noindent{\bf SU(5) GUT in 5D}: The first realistic
field theory model in which the gauge symmetry is broken by
orbifold boundary conditions is a 5D supersymmetric SU(5) GUT on $
M^4 \times S^1/(\Z_2\times \Z'_2)$ where $M^4$ is the 4D Minkowski
spacetime \cite{Kawamura}. The $\Z_2$ and
 $\Z_2'$ are as before. This compact space is shown in
Fig. \ref{s1z2}(b) with the fundamental region $[0,{\pi R\over
2}]$. There are two 4D walls (3-branes) placed at the fixed points
$y=0$ and $y={\pi R\over 2}$.

In 5D supersymmetric theory, there are two kinds of
supermultiplets, vector and hyper multiplets. We introduce a
vector-multiplet $V$\index{N=2 supermultiplet!vector-multiplet}
and two hyper-multiplets $H^s (s=1,2)$ in the 5D bulk. The vector
multiplet $V$ transforms as the adjoint representation ${\bf 24}$
of SU(5). The vector multiplet $V$ consists of a vector boson
$A_M$,\footnote{Note that we separated $A_\mu$ and $A_5$.} two
bispinors $\lambda_L^{i} (i=1,2)$, and a real scalar $\Sigma$. The
hyper-multiplets $H^{(s)}$ consist of two complex scalar fields
and two Dirac fermions $\psi^{(s)} =
(\psi^{(s)}_L,\psi^{(s)}_R)^T$, which are equivalent to four sets
of chiral supermultiplets: $H^{(1)} = \{H_5 \equiv
(H_1^{(1)},\psi_L^{(1)}), \hat{H}_{\bar{5}} =
(H_2^{(1)},\bar{\psi}_R^{(1)})\}$ and $H^{(2)} = \{\hat{H}_5
\equiv (H_1^{(2)},\psi_L^{(2)}), H_{\bar{5}} =
(H_2^{(2)},\bar{\psi}_R^{(2)})\}$. Out of two hyper-multiplets, we
can choose incomplete multiplets from each to form a complete
multiplet. For example, one can form a complete multiplet $\fiveb$
of SU(5) by picking up ${\threeb}(H_{\bar C})$ of SU(3) from the
hyper-multiplets $H_{\bar 5}$ and ${\two}(H_d)$ of SU(2) from
$\hat {H}_{\bar 5}$, and $\five$ of SU(5) by picking up $\three$
of SU(3) from the hyper-multiplets ${H}_{{5}}$ and ${\bf 2}(H_u)$
of SU(2) from $\hat H_{{5}}$. There are more possibilities. Let us
assume that our visible world is 4D located at $y=0$ and that
three families of quark and lepton chiral supermultiplets,
$3(\Phi_{\bar 5} + \Phi_{10})$, are located at this brane. Then,
unlike the KK modes of the bulk fields, matter fields contain no
massive states.

The gauge invariant action is given by
\begin{eqnarray}
&S = \int d^5x {\cal L}^{(5)}
    + {1\over 2}\int d^5x  \delta(y) {\cal L}^{(4)}
\end{eqnarray}
where
\begin{eqnarray}
{\cal L}^{(5)} &=& {\cal L}^{(5)}_{YM} + {\cal L}^{(5)}_{H},
\nonumber \\
{\cal L}^{(5)}_{YM} &=&\textstyle
   -{1\over 2} {\rm Tr} F^2_{MN} + {\rm Tr} |D_M\Sigma |^2
   + {\rm Tr} (i\bar{\lambda}_i \gamma^M D_M \lambda^i)
   - {\rm Tr} (\bar{\lambda}_i [\Sigma, \lambda^i ]),
\nonumber \\
{\cal L}^{(5)}_{H} &=&
   |D_M H_i^{(s)}|^2 + i\bar{\psi}_{(s)} \gamma^M D_M \psi^{(s)}\nonumber
   - (i\sqrt{2} g_{(5)} \bar{\psi}_{(s)} \lambda^i H_i^{(s)} + {\rm h.c.})
\nonumber \\
&&\textstyle- \bar{\psi}_{(s)} \Sigma \psi^{(s)} -
 H_{(s)}^{\dagger i} \Sigma^2 H_i^{(s)}
   - \frac12{g^2_{(5)}} \sum_{m,A} (H_{(s)}^{\dagger i}
   (\sigma^m)_i^j T^A H_j^{(s)})^2,
\nonumber \\
{\cal L}^{(4)} &\equiv&
   \sum_{\rm 3\ families} \int d^2\bar{\theta} d^2\theta
   \left( \Psi^\dagger_{\bar{5}} e^{2g_{(5)} V^A T^A} \Psi_{\bar{5}}+
    \Psi^\dagger_{\overline{10}} e^{2g_{(5)} V^A T^A}
    \Psi_{\overline{10}} \right)
\nonumber \\
&&+\sum_{\rm 3\ families} \int d^2 \theta \left(
    f_{U(5)} H_5 \Phi_{(10)} \Phi_{(10)}
    + \hat{f}_{U(5)} \hat{H}_5 \Phi_{(10)} \Phi_{(10)} \right.  \nonumber \\
&& \left. + f_{D(5)} H_{\bar{5}} \Phi_{(\bar{5})} \Phi_{(10)}
    + \hat{f}_{D(5)} \hat{H}_{\bar{5}} \Phi_{(\bar{5})} \Phi_{(10)} \right)
    + {\rm h.c.} \,,\nonumber
\end{eqnarray}
where $\lambda^i \equiv (\lambda^i_L,
\epsilon^{ij}\bar{\lambda}_{Lj})^T$, $D_M \equiv \partial
-ig_{(5)} A_M (x^\mu,y)$, $g_{(5)}$ is a 5D gauge coupling
constant, $\sigma^{m}$ are Pauli matrices, the $T^A$ are $SU(5)$
generators, $V^A T^A$ is an SU(5) vector supermultiplet.

The Lagrangian is required to be invariant under the
$\Z_2\times\Z_2^\prime$
transformation,\index{orbifold!$\Z_2\times\Z_2$ in 5D}
\begin{eqnarray}
\Z_2\ :\ \ A_\mu (x^\mu, y) &\longrightarrow &
   A_\mu (x^\mu, -y) = PA_\mu (x^\mu,y)P^{-1}, \nonumber \\
A_5 (x^\mu, y) &\longrightarrow &
   A_5 (x^\mu, -y) = -PA_5 (x^\mu,y)P^{-1}, \nonumber \\
\lambda_L^1 (x^\mu, y) &\longrightarrow &
   \lambda_L^1 (x^\mu, -y) = -P \lambda_L^1 (x^\mu,y)P^{-1},
   \nonumber \\
\lambda_L^2 (x^\mu, y) &\longrightarrow &
   \lambda_L^2 (x^\mu, -y) = P \lambda_L^2 (x^\mu,y)P^{-1},
   \nonumber \\
\Sigma (x^\mu, y) &\longrightarrow &
   \Sigma (x^\mu, -y) = -P \Sigma (x^\mu,y)P^{-1}, \nonumber \\
H_5 (x^\mu, y) &\longrightarrow &
   H_5 (x^\mu, -y) = PH_5 (x^\mu,y), \nonumber \\
\hat{H}_{\bar{5}} (x^\mu, y) &\longrightarrow &
   \hat{H}_{\bar{5}} (x^\mu, -y) = -P \hat{H}_{\bar{5}} (x^\mu,y),
    \nonumber \\
\hat{H}_5 (x^\mu, y) &\longrightarrow &
   \hat{H}_5 (x^\mu, -y) = -P \hat{H}_5 (x^\mu,y), \nonumber \\
H_{\bar{5}} (x^\mu, y) &\longrightarrow &
   H_{\bar{5}} (x^\mu, -y) = P H_{\bar{5}} (x^\mu,y)\label{kawabc}\\
   \nonumber\\
\Z_2^\prime\ :\ \ {\rm replace}\ \ y&\rightarrow& y^\prime\quad
{\rm and\ } P\rightarrow P^\prime\quad {\rm in\ Eq. }\
(\ref{kawabc}).\nonumber
\end{eqnarray}

The $5\times 5$ matrix $P$ acts on the gauge index space, and the
consistency under Lorentz transformation determines the overall
signs of the transformation of $A_\mu, A_5$ and $\Sigma$, viz. Eq.
(\ref{gsbc}) where $\Lambda={\rm diag.}(1, 1, 1, 1, -1)$. It also
requires that the overall signs of $\lambda_L^1$ and $\lambda_L^2$
are different. When we choose
\begin{eqnarray}
&&P={\rm diag}(-1,-1,-1,1,1)\label{P}\\
&&P'={\rm diag}(1,1,1,1,1),\label{Pprime}
\end{eqnarray}
the SU(5) gauge symmetry is broken down to that of the standard
model (SM) gauge group, ${\rm G_{SM}=SU(3)}\times $SU(2)$\times
$U(1). Note that the boundary conditions do not respect the SU(5)
symmetry, because not all the SU(5) generators $T^A
(A=1,2,...,24)$ do commute with $P,$
\begin{equation}\label{WPcomm}
P T^a P^{-1} = T^a, \quad\quad P T^{\hat{a}} P^{-1} =
-T^{\hat{a}}\,,
\end{equation}
where $T^a$ are generators for ${\rm G_{SM}}$ and $T^{\hat{a}}$
are the generators corresponding to the coset SU(5)/${\rm
G_{SM}}$. In Table \ref{t_su5}, we list the parity assignments and
the mass spectrum of the KK modes of the bulk fields. Each Higgs
multiplet in $H_5 (\hat{H}_{\bar{5}}, \hat{H}_{5},H_{\bar{5}} )$
is divided into the SU(3)-color triplet $H_C (\hat{H}_{\bar{C}},
\hat{H}_{C},H_{\bar{C}} )$ and the SU(2)-weak doublet
$H_u(\hat{H}_d, \hat{H}_u, H_d)$. Note that only $H_u$ and $H_d$
have zero modes. All the color triplet fields have masses of order
the KK scale, $\sim 1/R$. Thus the doublet-triplet splitting
problem of SU(5) is nicely resolved by assigning the boundary
conditions given in Eq.~(\ref{kawabc}).

The doublet-triplet splitting realized above is due to the moding
by $\Z_2$, i.e. with the orbifold conditions. In fact, this kind
of the orbifold doublet--triplet splitting was observed long time
ago in string orbifolds. But the threshold for understanding
string orbifold is much higher than the field theoretic orbifold.
In any case, the reason that string or field theoretic orbifolds
remove some unwanted fields from the low energy spectrum is due to
the boundary conditions which allow only specific fields at low
energy. The boundary conditions in field theoretic orbifold as
discussed above distinguish the KK modes by the quantum numbers of
the moding discrete group. And the boundary conditions in string
orbifold with the moding discrete group is required to satisfy the
modular invariance conditions, which is far more restrictive than
the conditions we discussed in the field theoretic orbifold.

Before discussing the SO(10) GUT, let us scrutinize the roles of
$\Z_2$ and $\Z_2^\prime$. As Eq. (\ref{WPcomm}) shows explicitly,
the moding group breaks the GUT group. For the purpose of breaking
the SU(5) group, we need just one discrete group such as $\Z_2$.
We note that then the colored field $H_C$ and $H_{\bar C}$ carry
$\Z_2$ parity + implying massless colored scalars. Thus, the
purpose of introducing another $\Z_2^\prime$ is to make colored
scalars carry a negative $\Z_2^\prime$ parity and hence not
allowing massless colored Higgs fields. Note that
$\Z_2\times\Z_2^\prime$ removes all the unwanted gauginos heavy.
Since $\lambda^1$ carries $\Z_2$ parity --, \N=2\ supersymmetry is
broken down to \N=1. Addition of $\Z_2^\prime$ removes
$\lambda^{2\hat a}$ also. So, three objectives are achieved by
$\Z_2\times\Z_2^\prime$ out of which two can be responsible for
each $\Z_2$ and the third result(SUSY breaking) is a bonus. In
fact, as soon as we introduce a $\Z_2$, the \N=2 supersymmetry is
already broken down to \N=1. This scenario works so nicely in 5D.
Note however that with one extra dimension these two $\Z_2$s are
the maximum number of discrete groups we can introduce. With more
internal space, there exist much more possibilities.

Another point to note is the symmetry behavior at the fixed point
$y=0$ and $\frac{\pi R}{2}$. The operator $P^\prime$ of Eq.
(\ref{Pprime}) is the identity in the SU(5) group space, and hence
it does not contribute to the gauge symmetry breaking. However,
the operator $P$ of Eq. (\ref{P}) does not leave the SU(5)
generators invariant and hence is the relevant one to see the
symmetry property around the fixed point. The matrix valued gauge
field $A_M(x,y)$ has the following expansion in terms of $y$
\begin{equation}
A_M(x,y)=A_M(x)+f_{MN}(x)y^M+O(y^2)\label{fMN}
\end{equation}
where $f_{MN}$ is simply a function of $x$. At a fixed point with
a finite $y$, $y=\frac{\pi R}{2}$, $A_M(x,y)$ is not invariant
under the change $y\rightarrow -y$, i.e. under $P$. But at the
fixed point $y=0$, $A_M(x,y)=A_M(x)$ and the transformation under
$P$ leaves the SU(5) symmetry intact. In field theoretic orbifold,
this is the general feature: at the origin $y=0$ the gauge
symmetry is unbroken. On the other hand, at the fixed point with a
finite $y$ the gauge symmetry is a broken one if $P$ breaks the
symmetry.

 The $5\times 5$ matrix $P$ acts on the gauge index
space, and the consistency under Lorentz transformation determines
the overall signs of the transformation of $A_\mu, A_5$ and
$\Sigma$, viz. Eq. (\ref{gsbc}) where $\Lambda={\rm diag.}(1, 1,
1, 1, -1)$. It also requires that the overall signs of
$\lambda_L^1$ and $\lambda_L^2$ are different. When we choose $P$
and  $P'$ as given in Eqs. (\ref{P},\ref{Pprime}), the SU(5) gauge
symmetry is broken down to that of the standard model (SM) gauge
group, ${\rm G_{SM}=SU(3)}\times $SU(2)$\times $U(1). Note that
the boundary conditions do not respect the SU(5) symmetry, because
not all the SU(5) generators $T^A (A=1,2,...,24)$ do commute with
$P,$
\begin{equation}\label{WPcomm}
P T^a P^{-1} = T^a, \quad\quad P T^{\hat{a}} P^{-1} =
-T^{\hat{a}}\,,
\end{equation}
where $T^a$ are generators for ${\rm G_{SM}}$ and $T^{\hat{a}}$
are the generators corresponding to the coset SU(5)/${\rm
G_{SM}}$. In Table \ref{tsu5}, we list the parity assignments and
the mass spectrum of the KK modes of the bulk fields. Each Higgs
multiplet in $H_5 (\hat{H}_{\bar{5}}, \hat{H}_{5},H_{\bar{5}} )$
is divided into the SU(3)-color triplet $H_C (\hat{H}_{\bar{C}},
\hat{H}_{C},H_{\bar{C}} )$ and the SU(2)-weak doublet
$H_u(\hat{H}_d, \hat{H}_u, H_d)$. Note that only $H_u$ and $H_d$
have zero modes. All the color triplet fields have masses of order
the KK scale, $\sim 1/R$. Thus the doublet-triplet splitting
problem of SU(5) is nicely resolved by assigning the boundary
conditions given in Eq.~(\ref{kawabc}).
\begin{table}[t]
\renewcommand{\arraystretch}{1.6}
\begin{center}
\begin{tabular}{|l|l|l|l|}
\hline\hline
4D fields & Quantum numbers & $\Z_2 \times \Z_2'$ & Mass \\
\hline $A_{\mu}^{a(2n)} $, $\lambda^{2a(2n)}$ & $({\bf 8}, {\bf
1}) + ({\bf 1}, {\bf 3})
 + ({\bf 1}, {\bf 1})$ & $(+, +)$ & $\displaystyle{{2n }}$ \\
$A_{\mu}^{\hat{a}(2n+1)}$, $\lambda^{2\hat{a}(2n+1)}$ &  $({\bf
3}, {\bf 2}) + (\bar{\bf 3}, {\bf 2})$
& $(+, -)$ & $\displaystyle{{2n+1 }}$ \\
\hline $A_{5}^{a(2n+2)}$, $\Sigma^{a(2n+2)}$, $\lambda^{1a(2n+2)}$
& $({\bf 8}, {\bf 1}) + ({\bf 1}, {\bf 3})
 + ({\bf 1}, {\bf 1})$ & $(-, -)$ & $\displaystyle{{2n+2 }}$ \\
$A_{5}^{\hat{a}(2n+1)}$, $\Sigma^{\hat{a}(2n+1)}$,
$\lambda^{1\hat{a}(2n+1)}$ & $({\bf 3}, {\bf 2}) + (\bar{\bf 3},
{\bf 2})$
& $(-, +)$ & $\displaystyle{{2n+1 }}$ \\
\hline $H_C^{(2n+1)}$ & $({\bf 3}, {\bf 1})$ & $(+, -)$ &
$\displaystyle{{2n+1
}}$ \\
$H_u^{(2n)}$ & $({\bf 1}, {\bf 2})$ & $(+, +)$ & $\displaystyle{{2n }}$ \\
$\hat{H}_{\bar{C}}^{(2n+1)}$ & $(\bar{{\bf 3}}, {\bf 1})$ & $(-,
+)$
& $\displaystyle{{2n+1 }}$ \\
$\hat{H}_d^{(2n+2)}$ & $({\bf 1}, {\bf 2})$ & $(-, -)$ &
$\displaystyle{{2n+2 }}$ \\
\hline $\hat{H}_C^{(2n+1)}$ & $({\bf 3}, {\bf 1})$ & $(-, +)$ &
$\displaystyle{{2n+1 }}$ \\
$\hat{H}_u^{(2n)}$ & $({\bf 1}, {\bf 2})$ & $(-, -)$ &
$\displaystyle{{2n+2
}}$ \\
$H_{\bar{C}}^{(2n+1)}$ & $(\bar{{\bf 3}}, {\bf 1})$ & $(+, -)$
& $\displaystyle{{2n+1 }}$ \\
${H}_d^{(2n)}$ & $({\bf 1}, {\bf 2})$ & $(+, +)$ & $\displaystyle{{2n}}$ \\
\hline
\end{tabular}
\end{center}
\label{tsu5} \caption{The $\Z_2 \times \Z_2'$ parities and the KK
masses, in units of $\frac{1}{R}$, of the orbifolded SU(5) bulk
fields. }
\end{table}

The doublet-triplet splitting realized above is due to the moding
by $\Z_2$, i.e. with the orbifold conditions. In fact, this kind
of the orbifold doublet--triplet splitting was observed long time
ago in string orbifolds. But the threshold for understanding
string orbifold is much higher than the field theoretic orbifold.
In any case, the reason that string or field theoretic orbifolds
remove some unwanted fields from the low energy spectrum is due to
the boundary conditions which allow only specific fields at low
energy. The boundary conditions in field theoretic orbifold as
discussed above distinguish the KK modes by the quantum numbers of
the moding discrete group. And the boundary conditions in string
orbifold with the moding discrete group is required to satisfy the
modular invariance conditions, which is far more restrictive than
the conditions we discussed in the field theoretic orbifold.

Before discussing the SO(10) GUT, let us scrutinize the roles of
$\Z_2$ and $\Z_2^\prime$. As Eq. (\ref{WPcomm}) shows explicitly,
the moding group breaks the GUT group. For the purpose of breaking
the SU(5) group, we need just one discrete group such as $\Z_2$.
We note that  the colored field $H_C$ and $H_{\bar C}$ carry
$\Z_2$ parity + implying massless colored scalars. Thus, the
purpose of introducing another $\Z_2^\prime$ is to make colored
scalars carry a negative $\Z_2^\prime$ parity and hence not
allowing massless colored Higgs fields. Note that
$\Z_2\times\Z_2^\prime$ removes all the unwanted gauginos heavy.
Since $\lambda^1$ carries $\Z_2$ parity --, \N=2\ supersymmetry is
broken down to \N=1. Addition of $\Z_2^\prime$ removes
$\lambda^{2\hat a}$ also. So, three objectives are achieved by
$\Z_2\times\Z_2^\prime$ out of which two can be responsible for
each $\Z_2$ and the third result(SUSY breaking) is a bonus. In
fact, as soon as we introduce a $\Z_2$, the \N=2 supersymmetry is
already broken down to \N=1. This scenario works so nicely in 5D.
Note however that with one extra dimension these two $\Z_2$s are
the maximum number of discrete groups we can introduce. With more
internal space, there exist much more possibilities.

Another point to note is the symmetry behavior at the fixed point
$y=0$ and $\frac{\pi R}{2}$. The operator $P^\prime$ of Eq.
(\ref{Pprime}) is the identity in the SU(5) group space, and hence
it does not contribute to the gauge symmetry breaking. However,
the operator $P$ of Eq. (\ref{P}) does not leave the SU(5)
generators invariant and hence is the relevant one to see the
symmetry property around the fixed point. The matrix valued gauge
field $A_M(x,y)$ has the following expansion in terms of $y$
\begin{equation}
A_M(x,y)=A_M(x)+f_{MN}(x)y^M+O(y^2)\label{fMN}
\end{equation}
where $f_{MN}$ is simply a function of $x$. At a fixed point with
a finite $y$, $y=\frac{\pi R}{2}$, $A_M(x,y)$ is not invariant
under the change $y\rightarrow -y$, i.e. under $P$. But at the
fixed point $y=0$, $A_M(x,y)=A_M(x)$ and the transformation under
$P$ leaves the SU(5) symmetry intact. In field theoretic orbifold,
this is the general feature: at the origin $y=0$ the gauge
symmetry is unbroken. On the other hand, at the fixed point with a
finite $y$ the gauge symmetry is a broken one if $P$ breaks the
symmetry.

The effective gauge symmetry below the KK compactification is the
common union of the symmetries at the fixed points(SU(5) and $\rm
G_{SM}$). In the above SU(5) example, the common union is the SM
gauge group $\rm G_{SM}$. For the bulk, one may consider the bulk
symmetry as SU(5) since the symmetry is broken only by the
boundary conditions. But, the massless gauge bosons in the bulk do
not form a complete SU(5) multiplet due to the boundary
conditions, and hence one can consider the bulk symmetry as $\rm
G_{SM}$. This example shows that one may find the effective 4D
gauge symmetry by studying the massless gauge bosons in the bulk
or by picking up the common union of the symmetries respected at
the fixed points.

\vskip 0.3cm \noindent{\bf SO(10) GUT in 6D}: In the preceding
subsection, we discussed the minimal GUT in 5D. The next simplest
GUT is SO(10) and in this subsection we discuss the SO(10) GUT in
6D. There are good references on 6D SO(10) \cite{HN_SO10,
ABC_SO10, Kyae_SO10}. One interesting feature of 6D SO(10) is that
each fixed point respects different gauge groups. In this case the
low energy effective theory is the common intersection of the
groups respected at each fixed point. This is the generalization
of the symmetry breaking we discussed in the previous subsection
and has more complex structure. This kind of the common
intersection as the gauge group appears in string orbifold also.

Group theoretically, the SO(10) GUT has some merits over the SU(5)
GUT except that not being the minimal one. Firstly, the fifteen
chiral fields are put in a single representation {\bf 16} together
with an SU(5) singlet neutrino and realizes our {\it theme of
unification}. Second, since it contains an SU(5) singlet neutrino
in the spinor {\bf 16} of SO(10),  it is possible to introduce
small Majorana neutrino masses through the see-saw mechanism. Of
course, one can introduce SU(5) singlets in the SU(5) GUT and
introduce a smiliar see-saw mechanism, but in the SO(10) GUT the
see-saw neutrino mass is related to other couplings dictated from
the SO(10) symmetry. Third, because the top and bottom quarks are
put in the same representation {\bf 16}, it is possible to relate
their masses, i.e. the so-called top-bottom unification is
possible. Thus, it seems that the SO(10) GUT has its own merit to
study.

In most orbifold models, the rank of the gauge group is not
reduced. Therefore, to break the SO(10) GUT down to the SM one
needs a Higgs mechanism.

A $\Z_2$ can break the gauge symmetry down to one of its maximal
subgroups. Thus, one discrete $\Z_2$ can break SU(5) down to its
maximal subgroup $\rm G_{SM}$. But the group SO(10) is bigger than
SU(5), and with one $\Z_2$ one cannot directly break SO(10) down
to $\rm G_{SM}$. As commented before, therefore, one cannot
achieve a proper SO(10) symmetry breaking in 5D, since in 5D the
discrete group on a circle can be at most $\Z_2\times\Z_2^\prime$.
Especially, if one $\Z_2$ is reserved for the doublet-triplet
splitting and breaking the \N=2 supersymmetry down to \N=1, then
one lacks enough directions for breaking SO(10) to $\rm G_{SM}$.
Thus, one needs more $\Z_2$ symmetries. Then, it is necessary to
go beyond 5D. The simple orbifolding we considered does not lower
the rank.

Since the rank of $G_{SM}$ is 4, a simple method will not work.
Actually, breaking the huge gauge group employs the symmetries of
the fixed points. In the previous field theoretic example of SU(5)
GUT, this method was briefly introduced. Also in string orbifolds
with Wilson lines, different fixed points respect different gauge
symmetries and a common union of the gauge symmetries respected by
twisted sectors(fixed points) is the gauge symmetry of the
effective 4D theory. Therefore, the field theoretic orbifolds and
string orbifolds have common features at some level. So, this
method can be used for breaking SO(10).

As noted above, orbifolding itself does not reduce the rank. We
need a Higgs mechanism to obtain the SM. To obtain
SU(3)$\times$SU(2) factor from SO(10) by orbifolding only before
considering the Higgs mechanism, we must work at least in 6D.
Before presenting a full 6D SO(10) GUT model, let us recapitulate
the group theory aspects of SO(10).

\vskip 0.3cm \noindent{\bf Subgroups of SO(10)}: The interesting
rank 5 subgroups of SO(10) are (i) the Georgi-Glashow(GG) group
SU(5) \cite{GG73} times U(1), (ii) the Pati-Salam(PS) group
SU(4)$_c\times $SU(2)$_L\times $SU(2)$_R$ \cite{Pati-Salam}, and
(iii) the flipped SU(5)(F-SU(5)), i.e. SU(5)$'\times$
U(1)$_X$\cite{Barr}. All these groups can be obtained when one
nontrivial $\Z_2$ boundary conditions are imposed. Since $\rm
G_{SM}$ is a subgroup of each Case (i), (ii), and (iii), $\rm
G_{SM}$ is a common union of SU(5)$\times$U(1), SU(4)$_c\times
$SU(2)$_L\times $SU(2)$_R$ and SU(5)$'\times$U(1)$_X$. In Table
\ref{Common}, the sixteen chiral fields are classified under these
three cases.
\begin{table}[t]
\begin{center}
\begin{tabular}{|l|l|l|l|}
\hline\hline
$\rm G_{SM}$ fields & GG-SU(5) & PS 422 & F-SU(5) \\
\hline $q $ & ${\bf 10}_F$ & ${\bf (4,2,1)}$ & ${\bf 10}_F$ \\
$l$ & $\overline{\bf 5}_F$
& ${\bf (4,2,1)}$ & $\overline{\bf 5}_F$\\
\hline
$e^+$ & ${\bf 10}_F$ & ${\bf (\bar 4,1,2)}$ & ${\bf 1}_F$\\
$u^c$ & ${\bf 10}_F$ & ${\bf (\bar 4,1,2)}$ & $\overline{\bf 5}_F$\\
$d^c$ & $\overline{\bf 5}_F$ & ${\bf (\bar 4,1,2)}$ & ${\bf 10}_F$\\
\hline
 $N$ & ${\bf 1}_F$ & ${\bf (\bar 4,1,2)}$ & ${\bf 10}_F$\\
\hline
\end{tabular}
\end{center}
\caption{The chiral fields are L--handed.}\label{Common}
\end{table}

Consider for example, the quark doublet $q$ and the lepton doublet
$l$. Both of these complete the PS representation $\bf (4,2,1)$
under the PS group. However, they belong to two different
representations under the GG SU(5). So, we must split {\bf 10} and
$\overline{\bf 5}_F$ so that $q$ and $l$ themselves become a
complete representation. It is most easily achieved from chopping
off $\overline{\bf 5}_F$ so that $l$ is split. Then, ${\bf 10}$ is
also split to produce $q$. When we chop off $\overline{\bf 5}_F$
into 3 + 2, the unbroken group is $\rm G_{SM}$. For the part of
the PS group, the fourth color is separated from the remaining
three colors to produce $q$ and $l$. This means that the common
intersection of the  SU(5)$_{\rm GG}\times $U(1) and the PS group
is $\rm G_{SM}\times U(1)$. In this way, one can confirm that the
common intersection of any two columns of Table contains $\rm
G_{SM}$.  If we considered the PS group and the F-SU(5) group,
then the common intersection is again ${\rm G_{SM}\times U(1)}$.
Similarly, the common intersection of SU(5)$_{\rm GG}\times$U(1)
and the F-SU(5) is $\rm G_{SM}\times$ U(1).

 The subgroup structure of the SO(10)
can be understood more clearly when we classify the 45 generators
$T^a$ of SO(10) The generators of SO(10) are represented by {\it
imaginary and antisymmetric $10 \times 10$ matrices}. A standard
way of representing these matrices is by embedding the U($N$)
group into the SO(2$N$) group. Then, it is convenient to write
these imaginary and antisymmetric generators as direct products of
$2 \times 2(\sigma_0,\sigma_a)$ and $5 \times 5(S,A)$ matrices,
giving
\begin{equation}
SO(10):\ \ \sigma_0 \otimes A_5, \ \sigma_1 \otimes A_5, \
\sigma_2 \otimes S_5,\ \sigma_3 \otimes A_5\ .\label{SO10}
\end{equation}
 Here $\sigma_0$ and $\sigma_{1,2,3}$ are the $2\times 2$ unit
matrix and the Pauli matrices; $S_5$ and $A_5$ are fifteen real
and symmetric $5 \times 5$  matrices, and ten imaginary and
antisymmetric $5 \times 5$ matrices, respectively. The U(5)
subgroup of SO(10) is then generated by
\begin{equation}
U(5):\ \ \sigma_0 \otimes A_5,\ \  \sigma_2 \otimes S_5\label{U5}
\end{equation}
whose total number is 25 the number of U(5) generators.

 By embedding the SM gauge group into this U(5), we can
divide the $5\times 5$ matrix further by choosing the first three
indices 1,2,3 for the SU(3)$_c$ and the last two indices 4,5 for
the SU(2)$_L$. Then, $A_3, S_3, A_2,$ and $ S_2$ contain
 the SM group generators.\footnote{
 $A_n$ and $S_n$ are $n\times n$ matrices.} The total number of
these are 13 out of which the identity generator is not belonging
to the SM gauge group. The remaining 12 generators are those of
the SM. Now, let us denote the left-over pieces of $A_5$ and $S_5$
as $A_X$ and $S_X$. Then, the generators of the Georgi-Glashow
SU(5)$_{\rm GG}\times$U(1) subgroup can be grouped as
\begin{equation}
\hspace{-1in}
  {\rm SU(5)_{GG} \times U(1)}:
\begin{array}{ccc}
  \underline{\sigma_0 \otimes A_3}, &
  \underline{\sigma_0 \otimes A_2}, & \sigma_0 \otimes A_X \\
  \underline{\sigma_2 \otimes S_3}, &
  \underline{\sigma_2 \otimes S_2}, & \sigma_2 \otimes S_X.
\end{array}\label{GGrep}
\end{equation}
The total number of generators in Eq. (\ref{GGrep}) is 25.

In a similar manner, we can choose the flipped SU(5) generators
\cite{flipped3,dkn3}. The relevant SO(10) subgroup is SU(5)$_{\rm
F} \times$U(1)$_X^\prime$ whose generators are
\begin{equation}
\hspace{-1in}
  {\rm SU(5)_F\times U(1)_X'}:
\begin{array}{ccc}
  \underline{\sigma_0 \otimes A_3}, &
  \underline{\sigma_0 \otimes A_2},
  & \sigma_1 \otimes A_X \\
  \underline{\sigma_2 \otimes S_3}, &
  \underline{\sigma_2 \otimes S_2},
  & \sigma_3 \otimes A_X.
\end{array}
\label{FSU5}
\end{equation}
Here also, the total number of generators is 25.

Finally, the generators of the Pati-Salam group SU(4)$_c
\times$SU(2)$_L \times $SU(2)$_R$ are given by
\begin{equation}
  {\rm PS\ 422}:
\begin{array}{cc}
  (\underline{\sigma_0},\sigma_1,\sigma_3) \underline{\otimes A_3},
  & \underline{\sigma_2 \otimes S_3 } \\
  (\underline{\sigma_0},\sigma_1,\sigma_3) \underline{\otimes A_2},
  & \underline{\sigma_2 \otimes S_2}.
\end{array}
\label{422gen}
\end{equation}
where the upper line lists the SU(4) generators and the lower line
lists the SU(2)$\times$SU(2) generators. Here the total number of
generators is 21.

\begin{figure}[h]
\begin{center}

\begin{picture}(400,210)(20,30)

\SetWidth{1.5} \Line(200,60)(153.8,140) \Line(200,60)(246.2,140)
\Line(153.8,140)(246.2,140) \Text(200,120)[c]{\bf G$_{\bf
SM}$}\Text(200,105)[c]{\bf $\times$U(1)}

\SetWidth{0.8}
\CArc(200,166.7)(53.33,-30,210)\Text(200,166)[c]{\Large
SU(5)$_{\rm GG}$}\Text(200,187)[c]{\Large U(1)$\times$}
\CArc(153.8,86.7)(53.33,90,330)\Text(145,80)[c]{\Large PS}
\CArc(246.2,86.7)(53.33,210,90)\Text(260,80)[c]{\Large F--SU(5)}

\end{picture}
\caption{The common intersection of SO(10) subgroups.}
\label{CommonI}
\end{center}
\end{figure}
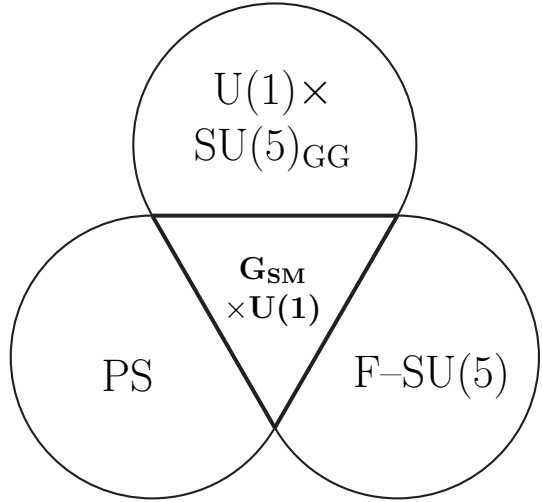

It can be clearly seen from the above decompositions that the
intersection of any combination of two of the GG-SU(5), F-SU(5)
and PS 422, is $\rm G_{SM}\times$ U(1), whose common generators
are
\begin{equation}
\begin{array}{cc}
  \sigma_0 \otimes A_3, & \sigma_0 \otimes A_2,  \\
  \sigma_2 \otimes S_3, & \sigma_2 \otimes S_2,
\end{array}
\end{equation}
which are underlined in Eqs. (\ref{GGrep}, \ref{FSU5},
\ref{422gen}). The common intersection of these SO(10) subgroups
is shown schematically in Fig. \ref{CommonI}. The U(1) generator
in the common intersection is $\sigma_2\otimes I_5$.

In 5D models, we noted that there are only two $\Z_2$ actions
available.\footnote{It was commented above that another $\Z_2$ is
categorically remembered for supersymmetry breaking and the
doublet-triplet splitting.}  If we use the second $\Z_2$ action to
break gauge group further, there remain unwanted massless fields
from $A_5$ components of the vector multiplet. See, for example,
Ref. \cite{Kyae_SO10}. These do not form a complete representation
of GUT groups. Thus the gauge coupling unification may not be
accomplished. To break SO(10) down to $\rm G_{SM}(\times$U(1))
just by orbifolding, we need to go at least to 6D. But we need a
Higgs mechanism to reduce the rank 5 of SO(10) down to the rank 4
SM. In this case, 5D SO(10) models can be considered even if one
$\Z_2$ is used for breaking the gauge group \cite{derm,kimraby}
where the rank preserving breaking
SO(10)$\rightarrow$SU(4)$\times$SU(2)$\times$SU(2) is achieved by
orbifolding and SU(4)$\times$SU(2)$\times$SU(2)$\rightarrow$SM is
realized by the Higgs mechanism $\langle (\bar 4,1,2)\rangle$.
Since we discussed 5D orbifolding already in terms of SU(5), let
us consider a 6D orbifolding.

\vskip 0.3cm \noindent{\bf 2D torus}: Now we proceed to discuss
compactifying two extra dimensions $x_5$ and $x_6$.\footnote{Note
the different convention for extra dimensions in string
compactification where extra dimensions are counted as
$4,\cdots,9$.} The 2D torus $T^2$ is defined by two vectors
$\hat{e}_1$ and $\hat{e}_2$ in the $\vec{y}$ plane $(y_1=x^5, y_2=
x^6)$. We identify two points $\vec{y}_A$ and $\vec{y}_B$ by the
following discrete translational symmetries,
$$
\vec{y}_B\equiv\vec{y}_A+m \hat{e}_1+n \hat{e}_2\ \quad {\rm for\
integers}\ \ m,\ n.
$$
Consider an SO(10) vector supermultiplet $V$ in the Hilbert space.
If the SO(10) is the symmetry on the torus, the SO(10) vector
supermultiplet must remain as an SO(10) vector supermultiplet
under the above discrete translation since the two points are
identified, but the components can be rearranged. Mathematically,
this is written as,
\begin{eqnarray}   \label{tor-bc-eq}
  V(\vec{y}+\vec{e}_i) = P_i V(\vec{y}) P_i^{-1}
\end{eqnarray}
where $P_i(i=1,2)$ leaves the SO(10) algebra invariant. On the
other hand, if $P_i$ does not commute with some SO(10) generators,
the SO(10) symmetry is broken. The interesting form of the $P_i$
matrices related with the three interesting subgroups of SO(10)
are\footnote{In Ref. \cite{Kyae_SO10}, $P_{\rm GG}, P_{\rm F}$ and
$P_{\rm PS}$ are represented as $P_2, P_3$ and $P_4$,
respectively.}
\begin{eqnarray} \label{pm-tor}
  P_{\rm GG}   &\equiv& \sigma_2 \otimes I_5,\label{GGP} \\
  P_{\rm F} &\equiv& \sigma_2 \otimes {\rm diag}(1,1,1,-1,-1),
  \label{FP} \\
  P_{\rm PS}  &\equiv& \sigma_0 \otimes {\rm diag}(1,1,1,-1,-1)\,.
\label{PSP}
\end{eqnarray}

It is easy to find out $P_{\rm GG}$ and $P_{\rm F}$. For these,
one needs to kill 20 generators among those in Eq. (\ref{SO10}).
The $\sigma_2$ in Eqs. (\ref{GGP}, \ref{FP}) does not commute with
the operators containing $\sigma_1$ and $\sigma_3$ in Eq.
(\ref{SO10}) and hence can exclude 20 generators. If the U(5) part
is the identity as given in Eq. (\ref{GGP}), then the traceless
$5\times 5$ matrices are the generators of SU(5). Since $Q_{\rm
em}$ is traceless in the GG model and in SO(10), $Q_{\rm em}$
should not contain the U(1) piece in U(5) and one concludes that
$P_{\rm GG}$ leaves SU(5)$_{\rm GG}\times$U(1) invariant.

If the U(5) part is not the identity as given in Eq. (\ref{FP}),
then traceless $Q_{\rm em}$ in SO(10) must contain the U(1)$_X$
piece in the flipped SU(5), which is the flipped SU(5). Here, the
commutators of $\sigma_2 \otimes {\rm diag}(1,1,1,-1,-1)$ with the
generators $\sigma_1\otimes A_X$ and $\sigma_3\otimes A_X$ are
non-vanishing but put again in the set. Note that there are two
non-vanishing factors $[\sigma_2,\sigma_{1,3}]$ and $[{\rm
diag}(1,1,1,-1,-1), A_X]$. If we assign one -- for a non-vanishing
commutator, we end up with + for the two commutator factors; thus
the F--SU(5) gauge bosons carry the + parity.

For the PS model, one must exclude 24 generators from Eq.
(\ref{SO10}). One can construct this number by $4\times 6$. Thus,
any operator containing $A_X$ and $S_X$ must be excluded, which is
possible with ${\rm diag}(1,1,1,-1,-1)$. But, the $\sigma$ part
must be intact and hence we obtain $P_{\rm PS}$ of Eq.
(\ref{PSP}).

These matrices leave the unbroken group generators even while the
others odd, under the transformation $T^a \longrightarrow P_i T^a
P_i^{-1}$. When the boundary condition Eq.(\ref{tor-bc-eq}) is
imposed, only the gauge fields for the corresponding subgroup can
have zero modes. Thus SO(10) is broken to the corresponding
subgroup. Observing the intersection of any two subgroups among
Georgi-Glashow, flipped-SU(5) or Pati-Salam leads to ${\rm
G_{SM}}\times$ U(1), choosing any two of (\ref{pm-tor}) will break
SO(10) gauge group down to that of ${\rm G_{SM}}\times$ U(1).
Using those $P$s in Eqs. (\ref{GGP}, \ref{FP}, \ref{PSP}) as the
boundary conditions, one can break the SO(10) gauge symmetry. But,
the simple torus compactification does not lead to chiral fermions
at low energy. To obtain chiral fermions, we must mod out by some
discrete group to obtain fixed points.

\vskip 0.3cm \noindent{\bf $\Z_2$ orbifold}: To obtain chiral
fermions, we must mod out the 2D torus by some discrete group to
build a two dimensional orbifold. The frequently used ones are
$T^2/\Z_2, T^2/\Z_3, T^2/\Z_6$ and $T^2/(\Z_2\times \Z_2')$.

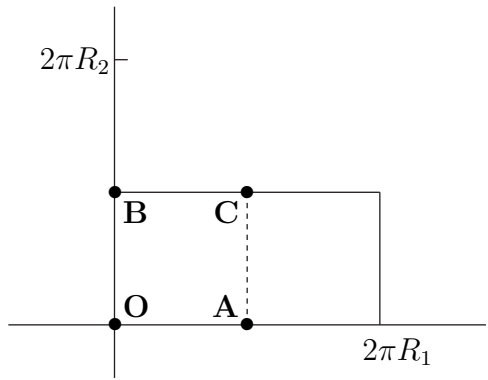
\begin{figure}[t]
\begin{center}

\begin{picture}(200,200)(-60,-30)

\Line(-90,0)(90,0) \Line(50,0)(50,50) \Line(50,50)(-50,50)
\Line(-50,-20)(-50,120) \DashLine(0,0)(0,50){2}
\Text(-47,3)[bl]{\bf {O}} \Text(-47,47)[tl]{\bf {B}}
\Text(-3,47)[tr]{\bf {C}} \Text(-3,3)[br]{\bf {A}}
\Text(-65,100)[c]{$2\pi R_2$} \Text(57,-5)[t]{$2\pi R_1$}
\Line(-50,100)(-45,100) \Text(-50,0)[c]{$\bullet$}
\Text(0,0)[c]{$\bullet$} \Text(0,50)[c]{$\bullet$}
\Text(-50,50)[c]{$\bullet$}
\end{picture}
\caption{The $T^2/\Z_2$ orbifold in the $(y_1,y_2)$ plane.
 The orbifold fixed points are denoted by bullets.  The
 fundamental region is the rectangle whose central line is the dashed line.}
\label{6dorbifig}
\end{center}
\end{figure}

In this section, we discuss a model based on $T^2/\Z_2$ orbifold.
This orbifold can be constructed by the following identifications:
\begin{eqnarray}
&&T_i : (x^\mu, \vec{y}) = (x^\mu, \vec{y}+\hat{e}_i) \label{idtorus} \\
&&\Z_2 \   : (x^\mu, \vec{y}) = (x^\mu, -\vec{y})\,,
\end{eqnarray}
where, with $R_1=R_2=R$,
\begin{equation}
\hat{e}_1 = (2\pi R, 0)\quad\quad \hat{e}_2 = (0, 2\pi R).
\end{equation}
Note that the $\Z_2$ action is $\pi$ rotation around the origin in
the $(y_1, y_2)$ plane, i.e. the reflection around the origin.
$T_i$ are the usual torus translations along the direction
$\hat{e}_i$. The fundamental region\index{fundamental region} can
be taken as the rectangle shown in Fig. \ref{6dorbifig}. The
orbifold fixed points are those that remain at the same point
under the $\pi$ rotation. Of course, here one uses the
identification (\ref{idtorus}) on the torus.  On the torus, there
are four orbifold fixed points at $\vec{y} = (0,0),\ (\pi R, 0),\
(0,\pi R),$ and  $(\pi R, \pi R)$. At a fixed point, a certain
gauge symmetry is respected as we have seen in the 5D case.

\vskip 0.3cm \noindent{\bf Gauge symmetries at fixed points}: Let
us proceed to discuss the gauge symmetries at fixed points. Here
in 6D also, different fixed points respect different gauge
symmetries. Of course, at the origin the gauge group is unbroken
as we have discussed in the previous subsection.

 The 6D \N=1 SO(10) gauge multiplet can be decomposed into
4D \N=1 SUSY multiplets: a vector multiplet $V$ and chiral adjoint
multiplet $\Phi$. The bulk action is given by

\begin{eqnarray}
  S &=& \int d^6 x \Biggl\{ {1 \over 4 k g^2}
    {\rm Tr} \left[ \int d^2\theta {\cal W}^{\alpha} {\cal W}_{\alpha}
    + {\rm h.c.} \right]
\label{so10action} \\
  &+& \int d^4\theta {1 \over k g^2}
    {\rm Tr} \Biggl[ (\sqrt{2} \partial^\dagger + \Phi^\dagger)
    e^{-V} (-\sqrt{2} \partial + \Phi) e^{V} +
    \partial^\dagger e^{-V} \partial e^V  \Biggr] \Biggr\},
\nonumber
\end{eqnarray}
where $V=V^a T^a$, $\Phi=\Phi^a T^a$, ${\rm Tr}\
[T^a,T^b]=k\delta^{ab}$ and $\partial=\partial_5 - i
\partial_6$.

Under the torus translations we identify
\begin{eqnarray}
  V(\vec{y}+\hat{e}_i) &=& T_i V(\vec{y}) T_i^{-1}, \\
  \Phi(\vec{y}+\hat{e}_i) &=& T_i \Phi(\vec{y}) T_i^{-1}.
\end{eqnarray}
The projection matrices are chosen as
\begin{equation}
T_1=P_{\rm GG},\ \  T_2=P_{\rm F}.
\end{equation}
Under the orbifolding $\Z_2$ ($\pi$ rotation), we identify
\begin{eqnarray}
  V(-\vec{y}) &=& Z V(\vec{y}) Z^{-1},\label{Z2gauge}\\
  \Phi(-\vec{y}) &=& -Z \Phi(\vec{y}) Z^{-1},\label{Z2scalar}
\end{eqnarray}
with $Z = \sigma_0 \otimes I_5$. Note that $Z$ is not belonging to
an SO(10) gauge generator, viz. Eq. (\ref{SO10}), since it is real
and symmetric. It belongs to a discrete group $\Z_2$, and is used
to break supersymmetry.

At the fixed points on the orbifold, certain gauge transformation
parameters are forced to vanish. Remember that the 5D $\Z_2$
example in Eq. (\ref{fMN}) at the fixed point $y=0$ leaves the
SU(5) symmetry intact. But at the fixed point $y=\pi R$, the
$\Z_2$ transformation leaves only the SM group $\rm G_{SM}$
invariant. Note that the analysis would be simplified in terms of
$y^\prime=\pi R-y$ at the fixed point $y=\pi R$.
This observation is also applicable in the present 6D case. The
matter contents and interactions located at the fixed points O, A,
B, and C of Fig. \ref{6dorbifig} respect different gauge
symmetries. At O, the full SO(10) gauge symmetry is respected.

Next consider the fixed point A, $\vec{y}=(\pi R,0)$. At this
fixed point, it is easier to analyze in terms of $y_1^\prime=\pi
R-y$ and $y_2$. Under $\Z_2$ and discrete translation along the
$\hat e_1$ direction, the vector multiplet transforms as
\begin{eqnarray}
\Z_2&:& V(\pi R,0)=V(-\pi R,0)\nonumber \\
T_1&:& V(\pi R,0)=T_1 V(-\pi R,0) T_1^{-1}\nonumber
\end{eqnarray}
where $T_1=P_{\rm GG}$. These two equations are consistent only if
$[T_1,V(\pi R,0)]$ = 0 at the fixed point A. We can check that the
SO(10) generators commuting with $T_1$ are those of SU(5)$_{\rm
GG}\times$U(1). Thus, the gauge symmetry at A is SU(5)$_{\rm
GG}\times$U(1). Namely, the wave function for every gauge field
outside SU(5)$_{\rm GG}\times $U(1) vanishes at the fixed point A.
Thus, interactions at this fixed point preserve only the
SU(5)$_{\rm GG} \times $U(1). Similarly, interactions at point B
where $T_2=P_{\rm F}$ need only preserve SU(5)$_{\rm F}\times
$U(1)$_X$.

At the fixed point C, one must apply both $\hat{e}_1$ and
$\hat{e}_2$ translations to compare it with the result from
performing the $\Z_2$ rotation. and hence one finds $V(\pi R, \pi
R))=T_2 T_1 V(-\pi R, -\pi R) T_1^{-1} T_2^{-1}$. From Eqs.
(\ref{GGP}) and (\ref{FP}), one notes that $T_2 T_1 = \sigma_0
\otimes {\rm diag}(1,1,1,-1,-1)\longrightarrow P_{\rm PS}$, which
commutes only with the generators listed in Eq.~(\ref{422gen}).
Thus, at the fixed point C the gauge symmetry is the PS group
SU(4)$_c \times $SU(2)$_L\times $SU(2)$_R$.

From Eqs. (\ref{Z2gauge}, \ref{Z2scalar}) one notes that the 4D
\N=1 supersymmetry is preserved at each fixed point.

We could have taken $T_1=P_{\rm PS}$ and  $T_2=P_{\rm GG}$. Then,
the fixed point A preserves SU(4)$_c \times $SU(2)$_L \times
$SU(2)$_R$, and the fixed point B preserves SU(5)$_{\rm GG}$
$\times $U(1).  The unbroken gauge group is SU(3)$_c \times
$SU(2)$_L \times $U(1)$_Y \times $U(1) as before.

Note that $P_{\rm PS}$ is a $T_{3R}$ rotation by angle $\pi$ since
$P_{\rm PS}$ has --1 entries in the $T_{3R}$ direction. Thus, the
relation $P_{\rm F}=P_{\rm PS}P_{\rm GG}$ shows that SU(5)$_{\rm
GG}$ is flipped by a $T_{3R}$.\footnote{Note that this
transformation is on the group elements themselves.} This implies
that the representations of SU(5)$_{\rm GG}$ are flipped to those
of the flipped SU(5):
\begin{eqnarray}
&& e^c_L\leftrightarrow N\nonumber\\
&& d^c_L\leftrightarrow u^c_L\,.\nonumber
\end{eqnarray}

\section{String orbifolds}

Now let us discuss string orbifolds. There exist a bosonic string
in 26D and five superstrings in 10D. Because there exist quarks
and leptons, we consider superstrings as possible scheme relevant
to nature. Here we will work with the \E88\ heterotic string
\cite{ghmrone}.

\subsection{Heterotic string}
 The heterotic string employs the idea of
the bosonic string and superstring. At the outset, it is not
obvious how two different space-times, 26D of bosonic string and
10D of NSR string, can be put in the same scheme. The breakthrough
comes from the observation that closed strings have two separate
sectors, the left(L)-moving and the right(R)-moving sectors, which
are the functions of $\tau+\sigma$ and $\tau-\sigma$,
respectively. So, for closed strings the Hilbert space can be a
direct product space of the L--moving and R--moving sectors.
Namely, the world-sheet operators $X^\mu$ and $\Psi^\mu$ can have
different closed string eigenstates for the L-- and R--sectors.
This heterotic string is shown that it has no tachyons and no
gravitational and gauge anomalies, and is finite to one loop.

The closed string physical Hilbert space is constructed as a
direct product space of L and R; $|{\rm
L-movers}\rangle\otimes|{\rm R-movers}\rangle.$ The heterotic
string uses this idea maximally.  We will stick to the so-called
NSR formalism. There is another called the GS string which has the
manifest spacetime supersymmetry.

The NSR string in 2D world sheet with the GSO projection has the
10D supersymmetry. The mode expansions of the R--movers are
\begin{eqnarray}
&& X^\mu_R=\frac12 x^\mu+\frac{i}{2}
p^\mu(\tau-\sigma)+\frac{i}{2}\sum_{n\ne 0}^\infty
\frac{\alpha_n^\mu}{n}
e^{-2in(\tau-\sigma)}\\
&&{\rm R:}\  \Psi^\mu_R=\sum_{n\in\Z}d_n^\mu e^{-2in(\tau-\sigma)}\\
&&{\rm NS:}\  \Psi^\mu_R=\sum_{r\in\Z+1/2}b_r^\mu
e^{-2in(\tau-\sigma)}
\end{eqnarray}
The mode expansions of the L--movers are\index{heterotic
string!mode expansion}
\begin{eqnarray}
X^\mu_L&=&
\frac12x^\mu+\frac12p^\mu(\tau+\sigma)+\frac{i}{2}\sum_{n\ne
0}^\infty \frac{\tilde\alpha_n^\mu}{n}
e^{-2in(\tau+\sigma)}\\
X^I_L&=& x^I+p^I(\tau+\sigma)+\frac12i\sum_{n\ne 0}^\infty
\frac{\tilde\alpha_n^I}{n} e^{-2in(\tau+\sigma)}.\label{XI}
\end{eqnarray}
The commutators of oscillators are
\begin{eqnarray}
&&[x^\mu,p^\nu]=-i\eta^{\mu\nu}\nonumber\\
&&[\alpha_n^\mu,\alpha_m^\nu]=[\tilde\alpha_n^\mu,\tilde\alpha_m^\nu]
=-n\delta_{n,-m}\eta^{\mu\nu}\nonumber\\
&&[\tilde\alpha_n^\mu,\alpha_m^\nu]=0\nonumber\\
&&[\tilde\alpha_n^I,\tilde\alpha_m^J]=n\delta_{n,-m}
\delta^{IJ}\label{ConstSpin}
\end{eqnarray}
\begin{equation}\label{CommIntxp}
[x^I,p^J]=\frac{i}{2}\delta^{IJ},
\end{equation}
and
\begin{equation}
\begin{array}{cc}
{\rm R:}\ & \{d_m^\mu,d_n^\nu\}=\{\tilde d_m^\mu, \tilde
d_n^\nu\}=-\delta_{m+n,0}\ \eta^{\mu\nu}\\
\\
{\rm NS:} & \{b_r^\mu,b_s^\nu\}=\{\tilde b_r^\mu, \tilde
b_s^\nu\}=-\delta_{r+s,0}\ \eta^{\mu\nu}
\end{array}
\end{equation}
with the anti-commutators for left and right movers of $\Psi^\mu$
vanishing. Of course, for $m=n=0$ the R sector gives the 10D Dirac
gamma matrices,
\begin{equation}
\{d_0^\mu, d_0^\nu\}=-\eta^{\mu\nu},
\end{equation}
enabling the R sector interpreted as the fermionic sector in
space-time.

\vskip 0.3cm \noindent{\bf Gravity sector}: Any closed string
theory contains a massless spin-2 field which is interpreted as
graviton. The existence of spin-2 graviton in closed string theory
is the basic argument for string theory to be the essence in the
unification of gravity with the other forces of elementary
particles. In our case, the graviton, dilaton, and the rank two
antisymmetric tensor fields are traceless symmetric, trace, and
antisymmetric parts of $\tilde\alpha_{-1}^\mu |0\rangle_\L
b_{-1/2}^\nu |0\rangle_\R$, respectively. There are also
superpartners of these fields from the R sector of right movers if
we replace the above right mover $b_{-1/2}^\nu |0\rangle_\R$ by
the Ramond states $|0,\pm\rangle_\R$ with $\pm$ being the $s_0$
components of ${\bf 8}\in |s_0\rangle\otimes|s_1\rangle\otimes
|s_2\rangle\otimes |s_3\rangle$. The choice of $s_0$ defines the
helicity. Remember that there are two right moving Ramond states
transforming as gauge group singlets, which correspond to two
helicities. Since the Fermi-Dirac statistics is inherited from the
right movers, the chiralities are also given by the Ramond sector
of the right movers. We define four dimensional chirality
operator\index{chirality!helicity in heterotic string}
$\gamma$\index{chirality!chirality operator} whose eigenvalue is
given by the product of components of spinorial ${\bf 8}$, $\gamma
= s_1 s_2 s_3 $.

\vskip 0.3cm \noindent{\bf Nonabelian gauge groups}: The heterotic
string idea, quantized as above, seems to be working at the
bosonic and superstring level. Since we introduced 26D for the
L--movers and 10D for the R--movers, it is better to work in a
common dimension. Thus, it is better to compactify the extra 16
dimensions of the L--moving bosons. These L--moving bosons will
supply with KK spectrum among which massless modes are of our
interest at low energy(low energy here means compared with the 10D
string scale). Our notation for dimensions is
\begin{equation}
M=1,2,\cdots,26,\ \ \ {\rm with\ 10\ interchangeable\ with\ 0.}
\end{equation}
Thus, the extra dimensions in the bosonic part $I$ corresponds to
$$
I\in \{11,12,\cdots,26\}
$$

Suppose we compactify one of the 16 dimensions, i.e. $I=26$ or the
26th dimension with a radius $R$, $ x^{26}\equiv x^{26}+2\pi Rn $
for any integer $n$. This compactification inserted into the modes
of (\ref{XI}) does not change $X^{26}$. Due to this torus
comactification of space-time, sometimes the string can wind
around the torus. For the mapping $x\rightarrow X$, let us use
$X^{26}=X_L^{26}+X_R^{26}$ even though we did not introduce $X_R$.
In terms of $X^{26}$, the inclusion of winding mode is given by
\begin{equation}
X^{26}(\tau,\sigma+\pi)=X^{26}(\tau,\sigma)+2\pi Rn.
\end{equation}
But there should be a constraint. The torus compactification, $
x^{26}\equiv x^{26}+2\pi Rn $, must give the identical result for
the winding mode also, i.e. $e^{ip^{26}x^{26}}$ must be a single
valued. This condition quantizes $p^{26}=p_L^{26}+p_R^{26}$
\begin{equation}
p^{26}=\frac{m}{R},\ \ \ m={\rm integer}.
\end{equation}
But we introduced only $X_L$ and we must set $x_R=p_R=0.$

\vskip 0.3cm \noindent{\bf Compactification}: With the above
example of the 1D torus compactification in mind, let us
compactify the space $x^I$ on a 16D torus. A 16D torus is defined
from a 16D lattice $\Lambda$\index{lattice!lattice $\Lambda$} with
the basis vectors $e_a^I\ (a=1,2,\cdots,16)$.\footnote{$I$ is for
space-time and $a$ is for the torus moduli space characterized by
16D lattice $\Lambda$.} Let us choose the length of these bases as
$\sqrt2$.

Borrowing the above $x^{26}$ compactification, we can identify
\begin{equation}\label{xcoord}
x^I\equiv x^I+ \sqrt2\pi \sum_{a=1}^{16}n_aR_ae^I_a
\end{equation}
where $R_a$ is the compactification radius in the lattice
direction $a$ and $n_a$ are integers. This compactification can
include string winding modes
\begin{equation}
X^I(\tau,\sigma+\pi)\equiv X^I(\tau,\sigma)+ \sqrt2\pi
\sum_{a=1}^{16}n_aR_ae^I_a=X^I(\tau,\sigma)+2\pi L^I
\end{equation}
where
\begin{equation}
L^I=\frac{1}{\sqrt2}\sum_{a=1}^{16}n_aR_ae^I_a.
\end{equation}
The mode expansion of $X^I(\tau,\sigma)$ can be written as
\begin{equation}\label{XLR}
X^I(\tau,\sigma)=x^I+p^I\tau+2L^I\sigma +\frac{i}{2}\sum_{n\ne 0}
\frac{1}{n}\left(\alpha_n^Ie^{-2in(\tau-\sigma)}+\tilde\alpha_n^I
e^{-2in(\tau+\sigma)} \right).
\end{equation}
Writing in terms of R-- and L--movers,
\begin{equation}\label{XRno}
X^I_R(\tau-\sigma)=x^I_R+p^I_R(\tau-\sigma) +\frac{i}{2}\sum_{n\ne
0} \frac{1}{n}\alpha_n^Ie^{-2in(\tau-\sigma)}
\end{equation}
\begin{equation}\label{XLyes}
X^I_L(\tau+\sigma)=x^I_L+p^I_L(\tau+\sigma) +\frac{i}{2}\sum_{n\ne
0} \frac{1}{n}\tilde\alpha_n^I e^{-2in(\tau+\sigma)},
\end{equation}
where we can identify $x^I=x_L^I+x_R^I$, and
\begin{eqnarray}
&p_R^I=\frac12(p^I-2L^I)\label{prI}\\
&p_L^I=\frac12(p^I+2L^I)
\end{eqnarray}
Since we have not introduced $X_R^I$, it is better to remove
$p_R^I$ which can be achieved by quantizing $p^I=2L^I$. Then, we
obtain $p_L^I=2L^I$. In addition, $x_R^I$ should be zero in Eq.
(\ref{XRno}). Thus, $x$ in (\ref{xcoord}) must be fully $x_L$,
\begin{equation}\label{xLcoord}
x^I_L\equiv x^I_L+ \sqrt2\pi \sum_{a=1}^{16}n_aR_ae^I_a.
\end{equation}
The quantization condition (\ref{CommIntxp}) shows that $2p_L^I$
is the $x_L$ translation generator. Thus, we require that $e^{2i
p_L^Ix_L^I}$ is single valued for a winding string rather than
$e^{i p_L^Ix_L^I}$ is.

Now it is convenient to define the dual\index{lattice!dual lattice
$\tilde\Lambda$} lattice $\tilde\Lambda$ which has the basis
vectors $e^{*I}_a$. Thus,
\begin{equation}
\sum_{I=1}^{16}e^I_a e^{*I}_b=\delta_{ab}.
\end{equation}
Then $p_L^I$ is given in the dual lattice by
\begin{equation}\label{pLI}
p_L^I= \frac{1}{\sqrt2}\sum_{a=1}^{16}\frac{m_a}{R_a}e^{*I}_a,\ \
\ m_a={\rm integer},
\end{equation}
so that
$$
2x_L\cdot p_L=2\pi n_aR_ae^I_a \frac{m_b}{R_b}e^{*I}_b=2\pi n_a
{m_a}
$$
and hence $2p_L$ is the translation operator for
$x_L$.\index{heterotic string!translation operator}

As in the bosonic and NSR string cases, let us choose the
light-cone gauge. The mass squared operator in the light-cone
gauge is
\begin{equation}
M^2=M_L^2+M_R^2,
\end{equation}
with the closed string condition $M_L^2=M_R^2$. For a mass
eigenstate, we have
\begin{equation}
M^2(|R\rangle\otimes|L\rangle)=(M_L^2|L\rangle)\otimes|R\rangle
+(M_R^2|R\rangle)\otimes|L\rangle=\lambda|R\rangle\otimes|L\rangle
\end{equation}
which can be true if $M_L^2=M_R^2=\frac{\lambda}{2}.$

For $M_R^2$ in superstring in 10D,
\begin{equation}\label{M2RH}
\frac14 M_R^2=\left\{\begin{array}{cc} \sum_{m=1}^\infty
(\alpha_{-m}^i\alpha_m^i+md_{-m}^id_{m}^i)&\ \ \ {\rm R\ sector}\\
\sum_{n=1}^\infty \alpha_{-n}^i\alpha_n^i+\sum_{r=\frac12}^\infty
rb_{-r}^ib_{r}^i-\frac12&\ \ \ {\rm NS\ sector}
\end{array} \right.
\end{equation}

\noindent For $M_L^2$, we use the 26D bosonic string compactified
on 16D torus,\index{heterotic string!mass formula}
\begin{equation}\label{M2LH}
\frac14 M_L^2=\frac12 \sum_{I=11}^{26}(p_L^I)^2+\tilde N-1
\end{equation}
where
\begin{equation}\label{Ntilde}
\tilde N=\sum_{n=1}^\infty (\tilde\alpha_{-n}^i\tilde\alpha_{n}^i+
\tilde\alpha_{-n}^I\tilde\alpha_{n}^I),
\end{equation}
and --1 in Eq. (\ref{M2LH}) is  $a=1$ in the closed bosonic string
mass.

\vskip 0.3cm \noindent{\bf Gauge symmetry enhancement}: It will be
instructive to see explicitly how the massless states come about.
Suppose we compactify one of the 16 dimensions, i.e. $I=26$ as
discussed above such that the resulting 25D bosonic theory is a
consistent string theory. It will fulfil our purpose since we want
to see just the effect of compactification. In field theory, we
expect one U(1) gauge boson from $g_{MN}.$ But in the present case
of string, we expect two U(1) gauge bosons from $g_{MN}$ and
$B_{MN}$. If massless winding modes are added, the extended gauge
group will be of rank 2. The difference from the previous
discussion is that we do not start from excluding the right mover
$p_R^I$, viz. Eq. (\ref{prI}), by $p^I=2L^I$.

Referring to Eqs. (\ref{XLR}), (\ref{XRno}) and (\ref{XLyes}), we
can write
\begin{eqnarray}
& p_R^{26}=\frac12 (p^{26}-2L)\label{pr26} \\
& p_L^{26}=\frac12 (p^{26}+2L)\label{pl26}
\end{eqnarray}
with $x^{26}=x^{26}_R+x^{26}_L$. Thus, from $M_L^2=M_R^2$, we have
$M^2=M_L^2+M_R^2=2M_R^2=2M_L^2.$ The 25D mass $M^2$ is
$p^Mp_M+(p^{26})^2$. Remembering this and referring to Eqs.
(\ref{M2LH}) and (\ref{Ntilde}), we obtain
\begin{eqnarray}
\frac18M^2&=&\frac14M^2_L=\frac12(p_L^{26})^2+\sum_{n=1}^\infty\tilde
\alpha_{-n}^{26}\tilde \alpha_{n}^{26}-\sum_{n=1}^\infty\tilde
\alpha_{-n}^{\mu}\tilde \alpha_{\mu n}
-1\nonumber\\
&=& \frac14M^2_R=\frac12(p_R^{26})^2
+\sum_{n=1}^\infty\alpha_{-n}^{26}
\alpha_{n}^{26}-\sum_{n=1}^\infty \alpha_{-n}^{\mu} \alpha_{\mu
n}-1.\label{MLMR}
\end{eqnarray}
Averaging over the above expressions, using Eqs. (\ref{pr26}) and
(\ref{pl26}), we obtain
\begin{eqnarray}
\frac18M^2&=&\frac18(p^{26})^2+\frac12
L^2+\frac12\sum_{n=1}^\infty\tilde \alpha_{-n}^{26}\tilde
\alpha_{n}^{26}-\frac12\sum_{n=1}^\infty\tilde
\alpha_{-n}^{\mu}\tilde \alpha_{\mu n}\nonumber\\
&&+\frac12 \sum_{n=1}^\infty\alpha_{-n}^{26}
\alpha_{n}^{26}-\frac12\sum_{n=1}^\infty \alpha_{-n}^{\mu}
\alpha_{\mu n}-1   \nonumber\\
&&\label{intmass}
\end{eqnarray}
Since $p^{26}$ is the translation generator, $e^{i2\pi p^{26}R}=1$
or $p^{26}=\frac{m}{R}$ for integer $m$. Using $M_L^2=M_R^2$ from
the expression given in (\ref{MLMR}), we obtain
\begin{eqnarray}
&\sum_{n=1}^\infty\alpha_{-n}^{26}
\alpha_{n}^{26}-\sum_{n=1}^\infty \alpha_{-n}^{\mu} \alpha_{\mu
n}-\sum_{n=1}^\infty\tilde \alpha_{-n}^{26}\tilde
\alpha_{n}^{26}+\sum_{n=1}^\infty\tilde \alpha_{-n}^{\mu}\tilde
\alpha_{\mu n}\nonumber\\
&=p^{26}L=mn^\prime\label{justbefore}
\end{eqnarray}
where both $m$ and $n^\prime=\frac{L}{R}$ are integers. Note that
(\ref{justbefore}) gives an integer eigenvalue because it is a sum
of number operators. Inserting these in the previous mass formula,
we obtain\footnote{We use $n$ instead of $n^\prime$.}
\begin{eqnarray}
\frac18M^2=\frac{m^2}{8R^2}+\frac{n^2R^2}{2}+\frac{N+\tilde N}{2}
-1
\end{eqnarray}
where $N-\tilde N$ is $mn$ of Eq. (\ref{justbefore}). So, it can
be rewritten as
\begin{equation}
\frac18M^2=\frac{m^2}{8R^2}+\frac{n^2R^2}{2}+\frac{mn}{2}+\tilde N
-1.
\end{equation}
For a critical radius $R=1/\sqrt2$, it becomes\index{heterotic
string!mass formula}
\begin{equation}
\frac18M^2=\frac{m^2+n^2}{4}+\frac{mn}{2}+\tilde N -1,
\end{equation}
which gives the massless solutions with $mn=1$ and $\tilde N=0$
and also $mn=-1$ and $\tilde N=1$. There are four solutions
$$
(m,n,\tilde N)= \ (1,1,0),\ (-1,-1,0),\ (1,-1,1),\ (-1,1,1)
$$
which provide four massless winding modes to have the gauge group
SU(2)$\times $SU(2) together with the aforementioned two KK modes.

As seen above the massless winding states need the special radius,
$R=\sqrt{\alpha' / 2}$, for U(1)$^2$ to be enhanced to
SU(2)$\times $SU(2). If we compactify the 16 internal spaces of
the left moving $X^I$, the needed {\it winding modes require the
same critical radius} for compactification, which means we look
for {\it only simply laced groups.}

\vskip 0.3cm \noindent{\bf Modular invariance}: To determine which
gauge groups string theory allows, we require a consistency
condition, known as the {\it modular invariance.} Consider a
one-loop diagram, which is a torus parametrized by
$\tau=\tau_1+i\tau_2$.\footnote{The modular parameter $\tau$
should not be confused with the world-sheet coordinate $\tau$.}
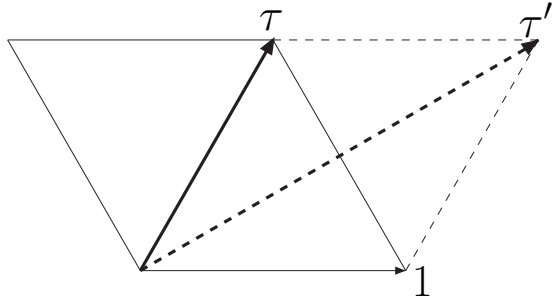
\begin{figure}[h]
\begin{center}
\begin{picture}(400,115)(0,-8)

\SetWidth{1.2}
\DashLine(100,0)(245,83.54){3}\ArrowLine(245,83.54)(250,87)
\Line(100,0)(150,87) \ArrowLine(148,83.54)(150,87)
\Text(150,95)[c]{\Large $\tau$}\Text(250,95)[c]{\Large
$\tau^\prime$}

\SetWidth{0.3} \Line(50,87)(150,87)\Line(200,0)(150,87)
\Line(100,0)(50,87) \Line(100,0)(198,0)\ArrowLine(196,0)(200,0)
\Text(207,-4)[c]{\Large 1}

\DashLine(200,0)(250,87){3}\DashLine(150,87)(250,87){3}
\end{picture}
\caption{$\tau$ defines a torus bounded by the lightly lined
parallelogram. By ${\cal T}$, it transforms to $\tau^\prime$ which
gives another parallelogram two sides of which are shown by
lightly dashed lines.}\label{torus}
\end{center}
\end{figure}

A torus can be represented by identifying the sides of a
parallelogram as shown in Fig \ref{torus}. Two edges of
parallelogram are described by two vectors 1 and $\tau$. The
string loop amplitude should not be affected by a
re-parametrization of the world-sheet which describes the
equivalent torus.

 Such equivalence, called the modular
transformation, generated by
\begin{equation}
 {\cal T}: \tau \to \tau+1, \quad {\cal S}: \tau \to -1/\tau.
\end{equation}

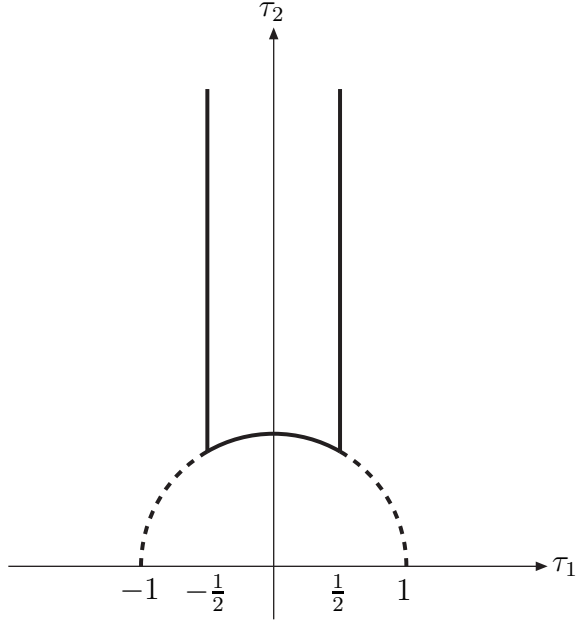
\begin{figure}[h]
\begin{center}
\begin{picture}(400,240)(0,-18)

\SetWidth{0.5} \Text(150,-9)[c]{$-1$}\Text(175,-10)[c]{$-\frac12$}
\Text(225,-10)[c]{$\frac12$}\Text(250,-9)[c]{$1$}

\Line(100,0)(300,0)\Line(200,-20)(200,200)
\ArrowLine(300,0)(302,0)\ArrowLine(200,200)(200,202)
\Text(306,0)[l]{$\tau_1$}\Text(200,210)[c]{$\tau_2$}

\SetWidth{1.5} \CArc(200,0)(50,60,120)
\DashCArc(200,0)(50,0,60){3} \DashCArc(200,0)(50,120,180){3}
\Line(225,43)(225,180)\Line(175,43)(175,180)

\end{picture}
\caption{The fundamental region is bounded by the bold lines and
arc.}\label{fundregion}
\end{center}
\end{figure}

This reduces the integration region to the {\it fundamental
region},
\begin{equation}
|\tau|>1,\quad  |\tau_1|<\frac12
\end{equation}
which is shown  in Fig. \ref{fundregion} and has a smaller size
than that of a point particle, so that the resulting loop integral
is free from an ultraviolet divergence. This means that its low
energy limit is anomaly free since anomaly comes from the failure
of regularization.

The modular invariance condition is contained in the metric of the
lattice. It is defined by $A^{IJ}=e^I \cdot e^J$, which is the
Cartan matrix:
\begin{enumerate}
\item Under $\cal T$, the amplitude is invariant only if
the lattice product $p_L \cdot p_L$ is even. It is shown that if
only the diagonal elements $A^{IJ}$ are even then the product is
even. This is the nature of Lie algebra diagrams, whose diagonal
elements of Cartan matrix are even.

\item Under $\cal S$, the lattice spanned by
$e^I$ is changed to its dual spanned by $e^{*I}$. For the
amplitude to be invariant under this transformation, there should
be no the change of volume, whose factor is given by $(\det
A)^{-1/2}$. Therefore, $\det A=1$ which is called self-dual
condition.
\end{enumerate}

It is known, from the theory of modular form,\index{modular
invariance!modular form} that there can be even, self-dual
Euclidian lattice only in $8k$ dimensions. Let us define the
lattice $\Gamma_{8k}$ as
\begin{equation}
 \textstyle \Gamma_{8k} = \{(n_1,n_2,\dots,n_{8k}),
 (n_1+\frac12,n_2+\frac12,\dots,n_{8k}+\frac12)|n_i \in \Z, \sum n_i
 \in 2\Z \} \label{gamma8}
\end{equation}
The former spans the root lattices\index{lattice!self-dual} of
SO($8k$) and the latter spans the spinorial representations of
SO($8k$). We know that E$_8$ roots can be made by the adjoint $\bf
120$ and the spinor $\bf 128$ of SO(16). They are vectors of
$\Gamma_8$, which is the only 8 dimensional even, self-dual
lattice.\index{lattice!SO(32), E$_8\times$E$_8$}\index{heterotic
string!$E_8\times E_8$}\index{heterotic string!Spin$(32)/{\bf
Z}_2$} The only lattices of dimension 16 are $\Gamma_8 \times
\Gamma_8$ and $\Gamma_{16}$, corresponding to E$_8\times$E$_8$ and
SO(32) root systems, respectively.\footnote{In fact for the case
of $\Gamma_{16}$, we have spinorial of SO(32) which is not the
adjoint. In this sense, we strictly call it a Spin(32)$/\Z_2$
group. However, we just use the name SO(32).} In 24 dimensions,
there are 24 lattices, the Niemeier lattices. When we relax the
Euclidian signature condition to include the Lorentz-like group
SO($m,n$), we can have a modular invariant lattice of the form
$\Gamma_a \times \Gamma_b $. This is the situation when we
compactify all the degrees of freedom except our 4D. The resulting
maximal possible group is SO(44)$\times$U(1)$^6$.

\vskip 0.3cm \noindent{\bf Massless states}: Finally, let us find
out the massless states. From (\ref{M2LH}), the lowest lying state
seems to be $p_L^2=0, \tilde N=0$ to have $\frac14M_L^2=-1.$
However, from (\ref{M2RH}) there is no right moving state with the
same mass. Thus, there is no tachyonic state. The lowest mass
states satisfying $M_L^2=M_R^2$ are the massless states where the
left movers are the following:
\begin{itemize}
\item $\tilde N=1$ and $p_L^2=0$: these states have the form
$\alpha^M_{-1}|0 \rangle_L$. In particular, those having group
space index $M=I$ provide U(1) generators, or the Cartan
subalgebra $H^I$.
\item $\tilde N=0$ and $p_L^2=2$: recall that $p_L$ is a 16
dimensional vector belonging to the lattice (\ref{gamma8}). They
are of the form $(\pm 1~\pm 1~0~\cdots~0)$ with permutations and
$(\pm \frac12 ~ \pm \frac12 ~ \cdots \pm \frac12)$ with even
numbers of minus sign. Indeed, they provide the root vectors of
E$_8\times$E$_8$ and SO(32) and the resulting low energy fields
are the charged generators $E^{p_L}$, or the ladder operators.
\end{itemize}
These vectors span the whole lattice (\ref{gamma8}). The R--movers
with $M_R^2=0$ are the NS states $b_{-\frac12}^\mu|0\rangle_R\
(\mu=1,2,\cdots,8)$ and the R states $d_0|0\rangle_R$(which is a
spinorial state). The NS states provide bosonic states and the R
states provide fermionic states since
$\{d_0^\mu,d_0^\nu\}=2\delta^{\mu\nu}$. The R--movers give
supersymmetric partners.

Thus, the direct product states $\alpha^\mu_{-1}|0\rangle_L\otimes
b_{-\frac12}^\nu|0\rangle_R$ are consistent $M^2=0$ states which
are the components of the graviton $g_{MN}$ and antisymmetric
tensor $B_{MN}$ in the light-cone gauge. The states
$\alpha^\mu_{-1}|0\rangle_L\otimes d_{0}|0\rangle_R$ are their
superpartners. In field theory limit, the single gravitino in 10D
gives the gravitational anomaly of --496. This should be canceled
by the following Majorana-Weyl spinors.

The direct product states $\alpha^I_{-1}|0\rangle_L\otimes
b_{-\frac12}^i|0\rangle_R$ are also consistent massless states.
They are spin-1 gauge bosons and $\alpha^I_{-1}|0\rangle_L\otimes
d_{0}^i|0\rangle_R$ are their superpartners, the gauginos. These
are only 16 gauginos, far short of 496. The 16 gauge bosons give
the gauge group U(1)$^{16}$ since the torus compactification of an
extra dimension gives a U(1) gauge group, the famous KK gauge
group. We need 480 more Majorana-Weyl fermions. Indeed, the
L--movers have more winding mode satisfying $M_L^2=0$ which is
possible if $\sum_{I=1}^{16}(p_L^I)^2=2.$ The number of these
winding states of the same length should be 480 for the theory to
be anomaly free. The $p_L^2$ expression gives the length in the
Hilbert space. As discussed, the winding mode is a 16 dimentional
vector in the dual lattice $\tilde\Lambda$, in the basis $\tilde
e_a$.

In the root space, any root can be expressed in terms of simple
roots. For the SU(3) example, simple roots are $\tilde e_1$ and
$\tilde e_2$, and non-simple roots are $\tilde e^\prime=\tilde
e_1+\tilde e_2, -\tilde e_1, -\tilde e_2$ and $-\tilde e^\prime$.
As in this SU(3) example, our discrete momenta in the 16
dimensional dual lattice $\tilde\Lambda$ are expressible in terms
of simple roots, $\tilde e_a\ (a=1,2,\cdots,16)$. But the lengths
of these nonzero roots are the same, as shown in (\ref{pLI})
expressed in another basis $e^{*I}_a$. The algebras allowing the
same-length nonzero roots are $A_n, D_n, E_6, E_7,$ and $E_8$.
Among these, the number of nonzero roots of rank 16 algebra must
be 480 to cancel the 10D gravitational anomaly, whether we take
simple or semi-simple algebra. It is possible only with algebras
$D_{16}$ and E$_8\times$E$_8^\prime$. The direct product states
$|p_L^I\rangle_L\otimes |b^i_{-\frac12}\rangle_R$ are spin-1, and
the equal length $p_L^I$'s together with the 16 KK gauge boson
generators provide all the generators for the gauge group. Thus,
heterotic string allows only two possibilities: SO(32) and
E$_8\times $E$_8^\prime$.

We can make a consistency check. One is to use the anomaly freedom
in 10D supergravity theory. This gives SO(32) and E$_8\times$E$_8$
as only possible gauge groups. In 10D \N=1 supergravity, there is
the spin-$\frac32$ gravitino. It has a gravitational anomaly of
unit --496. It can be cancelled by introducing 496 chiral
fermions. This miraculous cancellation can be tracked back to the
modular invariance.

To sum up, the low energy theory has the NS--NS fields, graviton
$g_{\mu \nu}$, dilaton $\phi$ and rank--2 antisymmetric tensor
field $B_{\mu \nu}$, in addition to E$_8\times$E$_8^\prime$ or
SO(32) gauge bosons. With the superpartners, we have ten
dimensional ${\cal N}=4$ supergravity,\footnote{Note that N=1 in
10D but \N=4 in 4D.} coupled with Yang--Mills gauge fields
E$_8\times$E$_8^\prime$ or SO(32).

\section{ Four dimensional strings on orbifolds}

Now let us construct 4D string models by compactifying the \E88\
heterotic string on orbifolds. There are two kinds of twisting,
one is the discrete action in the internal 6D and the other is to
accommodate the discrete operation in the group space. The former
is denoted as $v$ and the latter as $V$. By constructing 4D
models, one may succeed in obtaining
\begin{itemize}
\item Possibility of the SM gauge group
\item Three families
\item \sw0=$\frac38$
\item Doublet-triplet splitting
\end{itemize}

\noindent{\bf Strings on orbifolds}: An orbifold \cite{dhvw} is
obtained from a manifold by identifying points under a discrete
symmetry group ${\sf S}$. An element of this group $(\theta,v)$
transforms a point $x$ in ${\bf R}^n$ as,
\begin{equation}
x^i \to {(\theta,v)^i}_j x^j = {\theta^i}_j x^j + v^i.
\label{orbiaction}
\end{equation}
We are familiar with such moding. The torus $T^n$ is obtained by
moding out ${\bf R}^n$ by pure translation $(1,v)$.

The point group ${\sf P}$ is the subgroup of ${\sf S}$ with
automorphism $\theta$; just think of pure rotation. In addition,
we want to have such rotation compatible with the torus: we
require the action up to translation. Then, for every $\theta$
there is a unique vector $v$ such that $(\theta,v)$ is an element
in ${\sf S}$, up to translation $\Lambda$. To see it, compare two
elements with the same $\theta$,
\begin{equation}
(\theta,v)(\theta,u)^{-1}=(\theta,v)(\theta^{-1},
-\theta^{-1}u)=(1,v-u).
 \label{Pelement}
\end{equation}
Now $v-u$ should belong to $\Lambda$ otherwise those two are not
the symmetry in ${\sf S}$. Thus we will label the action by
$\theta$ only. This generalized point group $\bar {\sf P}$ is
identified with ${\sf S} / \Lambda$ and commute with $\Lambda$.
Thus we have another definition for the orbifold
\begin{equation}
{\bf R}^n / {\sf S} = {\bf R}^n / (\Lambda \times \bar{\sf P}) =
T^n / \bar {\sf P}. \label{orbifold}
\end{equation}
Eq. (\ref{orbifold}) shows that the torus is moded by the group
$\bar{\sf P}$, which is our main interest here. A set of images
due to such finite operations is called {\it orbit}, hence the
name {\it orbifold.}

From now on, let us confine our discussion on the discrete action
to $\Z_N$. With the $\Z_N$ symmetry,  $N$ successive operations
become identity and the number $N$ is called order. We
conventionally pair the coordinates
$$Z_i \equiv \frac1{\sqrt2} (X_{2i} + i X_{2i+1}),
\quad Z_{\bar \imath} \equiv
\overline{Z_i} = \frac1{\sqrt2}(X_{2i} - i X_{2i+1}).$$ The action
$\theta$ in $\bar {\sf P}$ is parameterized by its eigenvalues
$\phi_i$ as
\begin{equation}
Z_i \to  e^{ 2 \pi i \phi_i } Z_i. \label{phidef}
\end{equation}
Then, we get the orbifold (\ref{orbifold}) by the identification
under
$$ Z_i \sim  e^{ 2 \pi i \phi_i } Z_i, \quad Z_i \sim Z_i + e_i,
    \quad Z_{\bar \imath} \sim Z_{\bar \imath} + e_{\bar \imath}$$
with the vector $e_i$ defining the torus. We do the same action to
fermionic coordinate.

In general the action is not freely acting, that is, an orbifold
has singular points. Nevertheless, there exists a well-defined
description of strings on orbifold. Consider a point on the
lattice site, denoted by a vector $x^i e_i$. Here, $e_i$ is a unit
vector so that every entry of $x$ is an integer. By the action of
$\theta \in \bar {\sf P}$, the image $(\theta x)^i e_i$ is another
unit vector thus has also integral entries. This implies that
$\theta$ can be represented by a matrix with integer elements.
Thus, it is guaranteed that the trace and the determinant are also
integers,
\begin{equation}
 \chi = \det (1-\theta)=\prod 4 \sin^2 (\pi \phi_i),
 \label{numberfixedpt}
\end{equation}
with the index $i$ runs over compact dimensions. Then, $\chi$ is
the number of fixed points by Lefschetz fixed point theorem. If
there are fixed tori than points, then $\chi=0$. Here, we are
interested in compactifying (less than) six dimensions to obtain
our four dimensional(4D) world. Possible orbifolds are listed in
\cite{dhvw}.

\vskip 0.3cm \noindent{\bf Twisted string}: Under orbifold action,
the part of the string Hilbert space is invariant. This is the
{\it untwisted string}\index{untwisted string}. However we require
the {\it twisted string}\index{twisted string} in addition: it is
closed modulo ${\sf S}$ since the space is identified under ${\sf
S}$. We need it because of modular invariance, again. A twisted
string in general obtains a phase when we go along the string,
\begin{equation}
Z^i (\sigma+ \pi) = \theta Z^i (\sigma) =e^{2\pi i /N}  Z^i
(\sigma). \label{twistedsector}
\end{equation}
A successive $k(=1,\dots,N-1)$ twist define the $k^{{\rm th}}$
twisted sector. The consistency comes from the modular invariance,
as we will shortly see.

Take an example of the $T^2/\Z_N$ orbifold by which the two
dimensional torus lattice vectors are identified under $2 \pi /N$
rotation. The following mode expansion have the phase $e^{2\pi i k
/N}$
\begin{eqnarray}
 & Z^i(\sigma,\tau) = Z^i_{\rm f} + i \sqrt{\alpha' \over 8}
  \sum_{m \in {\bf Z}}
  \Big( {\tilde \alpha^i_{m+k/N} \over m+k/N}
         e^{-2i(m+k/N)(\tau+\sigma)}
         \nonumber\\
&      + {\alpha^i_{m-k/N} \over m-k/N}
         e^{-2i(m-k/N)(\tau-\sigma)} \Big).\label{Ziexp}
\end{eqnarray}
We can verify that, by complex conjugation, $(N-k)^{\rm th}$
sector has the same mode expansion as $k^{\rm th}$ sector with the
the two terms interchanged. Here, $Z_{\rm f}$ is the fixed point
under the space action obtained by $Z_{f}(\pi)=Z_{f}(0)$ in
(\ref{twistedsector}),
$$ Z_{f} = (1-\theta)^{-1} v, \quad v\in \Lambda. $$
The twisted string does not have a momentum nor a winding mode.
Recall that the relation (\ref{Pelement}) means that every sector
can be labelled by $v$ as well as $\theta$. Therefore, the set of
fixed points is isomorphic to $\Lambda / (1-\theta) \Lambda$.

The commutation relations of twisted oscillators are
\begin{eqnarray}
 &[ \tilde \alpha^i_{m+k/N},
  \tilde \alpha^j_{-n-k/N}
  ]=(m+\frac{k}{N})\delta^{ij}\delta_{mn},\\
&[\alpha^i_{m-k/N},
 \alpha^j_{-n+k/N}
  ]=(m-\frac{k}{N})\delta^{ij}\delta_{mn}.
\end{eqnarray}
Similarly, we obtain the mode expansions of the fermionic degrees
of freedom. The anticommutators of the oscillators are modified by
$\pm\frac{k}{N}$.

Since string tension is proportional to the length stretched, the
massless string state can only exist on the fixed point $X_{\rm
f}$. We see this from the mass shell condition of the original
heterotic string,
\begin{align}
 \textstyle {\frac14 \alpha' M^2_\L } &=\textstyle
  {\frac12 P^2 } + \tilde \Nr + \tilde c,
 \label{leftmassshell} \\
 \textstyle{\frac14 \alpha' M^2_{{\rm R}}} &=\textstyle
  \Nr + c \\
  M_\L^2 &= M_{{\rm R}}^2 =  M^2 /2 .
\end{align}
The $P^I \equiv p_L^I$ is quantized, sixteen dimensional vector.
We are forced to have the critical radius $R=\sqrt{{\alpha'}/2}$
because we have no corresponding right movers $p_R=0$.

Here, $\tilde \Nr$ is the oscillator number operator of L-movers,
$$
\tilde \Nr=\sum^8_m \sum_{n=1}^\infty \tilde \alpha^m_{-n-\phi^m}
\tilde \alpha^m_{n+\phi^m} + \sum^{16}_I \sum_{n=1}^\infty \tilde
\alpha^I_{-n} \tilde \alpha^I_n
$$
with twisting $\phi^i$. In the above expansion $\phi^i=1/N$. Then,
$\tilde c$ is the resulting zero point energy, from analytic
continuation of Riemann zeta function,
\begin{equation}
 f(\eta) = \frac12 \sum_{n=1}^\infty  (n+\eta) = -\frac1{24}
 + \frac14 \eta(1-\eta),
 \label{zeroptE}
\end{equation}
for each degree of freedom with $0 \le \eta \le 1$. The bosonic
and fermionic states have positive and negative contributions to
the zero point energy, respectively. From now on we will set
$\alpha'=1$ without confusion. We have seen that the untwisted
string has $\tilde c=-1$ for the left mover. For the right moving
superstring, $c=-\frac12,0$ for Neveu-Schwarz(NS) and
Ramond(R)states, respectively.

\vskip 0.3cm \noindent{\bf Breaking the gauge group}: To now we
have seen real space part of string. We embed the space group
action into an action in the group space by associating them
together. This will provide boundary condition to break group. We
saw this in FTO.

\vskip 0.3cm \noindent{\bf Embedding into gauge group}: Recall
that in the bosonic construction, group degrees of freedom is
described by extra sixteen coordinate compactified on tori. The
charge is represented by a momentum vector $P$. We associate the
orbifold action $\theta$ with the action
\begin{equation}
 |P\rangle \to \exp (2 \pi i V\cdot P) |P \rangle.
 \label{shift}\index{shift vector}
\end{equation}
The vector $V$ is called the {\it shift vector} and satisfies that
$N V$ belongs to the weight lattice for that to have the same
order as orbifolding. Every inner automorphism\index{automorphism}
can be represented by a shift vector. We use the weight vector $P$
and the state $|P\rangle$ interchangeably.

The unbroken gauge boson should not carry such phase
\begin{equation}
P \cdot V = \mbox{integer}. \label{master}
\end{equation}
To have a definite order we need both for the space-time and the
group space,
\begin{equation}
 \sum_i N \phi_i = \sum_I N V_I = 0 \quad (\rm{mod}~2)
 \label{mod2}
\end{equation}
where the modulo 2 condition is reserved for spinorial states,
e.g. in the $E_8\times E_8$ theory. In (spin $1/2$) fermionic
description the rotation by $2\pi$ is minus identity. Thus we
should be careful about mentioning the order.

\vskip 0.3cm \noindent{\bf Modular invariance}: Recall that the
modular invariance constrains the group lattice to
E$_8\times$E$_8$ or SO(32). The spacetime part is {\it
independently} modular invariant. By orbifolding, each part loses
modular invariance, however we can choose a special combination to
make the whole theory invariant. The modular transformation has
two generators
\begin{equation}
 {\cal T}: \tau \to \tau+1, \quad {\cal S}: \tau \to -1/\tau.
\end{equation}

The ${\cal T}$ generate the shift in the direction $\sigma$ whose
generator is world sheet momentum $P=\tilde L_0-L_0$\index{world
sheet momentum}. Since we have no preferred origin, these levels
should still match modulo integer\footnote{It is for the abelian
orbifold case. In nonabelian orbifold we have a much more
complicated condition.}
\begin{equation}
 \tilde L_0 - L_0 = \mbox{integer.} \label{matching}
\end{equation}
To see this, with a shift vector $V$, consider the left movers,
$$ \tilde L_0 = \sum \tilde \Nr_i
+ {(P+V)^2 \over 2 } + \tilde c
$$
where $\tilde\Nr_i$ is the number of oscillators $\tilde\alpha_i$
which now can assume a fractional number, and $\tilde c$ is the
zero point energy
$$
 \tilde c= 2 \sum_{i=1}^4 f(\phi_i) + \sum_{I=1}^{16} f(0).
$$
By the definition (\ref{phidef}) we rotated pairwise. In a similar
manner, we obtain for the right movers,
$$ L_0 = \sum \Nr_i  + 2\sum_{i=1}^4 f(\phi_i) $$
for the R sector. For the NS sector we shift $\phi$ by $1/2$ due
to minus sign. Note that the oscillator number $\Nr$ is a multiple
of $1/N$, and the zero point energy and $(P+V)^2$ are multiples of
$1/N^2$.  Thus, a necessary condition is imposed to make the
$1/N^2$ dependence vanish,
$$
\frac12 \sum_i \phi_i - \frac12 N \sum_i \phi_i^2 + \frac12 N
\sum_I (P_I+V_I)^2 = \mbox{integer}.
$$
Then, it follows that $N^2 (P^2 + 2 P\cdot V + V^2 - \phi^2) = 0$
modulo $2N$. Since we are in the even and self dual lattice, it
leads to
$$
P\cdot V=\frac{1}{N}
$$

\vskip 0.3cm \noindent{\bf Untwisted matter from $Z_3$ orbifold}:
The massless fields from bulk are given by
\begin{equation}
\begin{array}{c|c|c}
{\rm multiplicity}&{\rm condition}&{\rm fields}\\
\hline
 1 &P\cdot V
={\rm integer}&{\rm gauge\ boson}\\
3 & P\cdot V=\frac{1}{N}&{\rm  matter}
\end{array}
\end{equation}

\vskip 0.3cm \noindent{\bf Twisted matter from $Z_3$ orbifold}: In
calculating one loop amplitude we should consider boundary
conditions in both directions of the torus. Namely, along with
(\ref{twistedsector})
\begin{eqnarray}
& Z(\tau,\sigma+\pi) = h Z(\tau,\sigma), \\
& Z(\tau+\pi \tau_1,\sigma+\pi \tau_2) = g Z(\tau,\sigma).
\end{eqnarray}
The $g,h$ are elements of point group $\sf P$. We saw that under
$\cal S$ the lattice size is multiplied by the square root of the
determinant of Cartan matrix, $(\det A)^{-1/2}$, which is 1 for
self-dual lattices. In addition, it interchanges the boundary
conditions of the two worldsheet coordinates
$$ {\cal S}: (g,h) \to (h^{-1},g). $$
Recall that we defined the twisted sector by the nontrivial
boundary condition of the $\sigma$ coordinate. The part of the
untwisted sector $(h,1)$ goes into the twisted sector $(1,h)$.
Therefore, the twisted sector is also needed to close the modular
transformation.

\subsection{$\Z_3$ orbifold}

In this subsection, we present the $\Z_3$ orbifold example. It is
specified by a vector (\ref{phidef}),
$$
  \textstyle \phi = (\frac23 ~ \frac13 ~ \frac13 ).
$$
leading simultaneous rotations by $( \frac{4\pi}3, \frac{2\pi}
3,\frac{2\pi} 3)$ on each $T^2$. The rotations are commutative
because rotation generators are the Cartan generators of $SO(6)$.
Namely, it is an Abelian orbifold.

The lattice in one torus is depicted in Fig. The unit vectors are
the root vectors of $SU(3)$, hence it is called $SU(3)$ lattice.
All the points outside the {\it fundamental region} can be
transformed into points in the fundamental region, under some
rotations and translations. The half of the fundamental region is
surrounded by the fixed points. Each torus has three fixed points,
and hence there are 27 fixed points in total, shown schematically
in Fig.

We choose {\it standard embedding}, that is, the orbifold action
is associated with the shift in the root space
$$ V = \phi $$
with other degrees are not touched. Namely, the shift vector is
specified by
$$
  \textstyle V = (\frac 23 ~ \frac13 ~ \frac13
  ~0~0~0~0~0)(0~0~0~0~0~0~0~0).
$$
Sometimes repeating zeroes are abbreviatied using superscript. In
this, the above vector may simply written as
$$ V = \textstyle (\frac 23~\frac13^2~0^5)(0^8).$$
This automatically satisfies modular invariance condition. This is
the way that we related spin connection and gauge connection in
the Calabi-Yau (CY) compactification, so it turns out that the
standard embedding shares many feature with CY compactification.

\vskip 0.3cm\noindent{\bf Untwisted sector}: The original
$E_8\times E_8$ theory has mass shell condition
\begin{eqnarray}
  && {M_\L^2 \over 4} = {P^2 \over 2} +  \tilde \Nr - 1  = 0 \\
  && {M_\R^2 } = M_{\L}^2,
\end{eqnarray}
where $P$ is momentum vector in the internal 16 dimension. The
massless modes are oscillator state $\tilde\Nr = 1$ and
momentum-winding state $|P\rangle$ with $P^2 = 2$. The former
modes $\tilde \alpha_{-1}^I | 0 \rangle, I=1,\dots,16$ will be the
sixteen Cartan generators. They form the adjoint representations
of E$_8\times$E$_8$.

The untwisted states are a part of the original states invariant
under the $\Z_3$ action. They are of two kinds:
\begin{enumerate}
\item[(1)] each left and right mover itself is invariant
\item[(2)] those only invariant when combined with covariant right
movers
\end{enumerate}
Thus, the orbifold action (with (2)) breaks gauge group.

\vskip 0.3cm\noindent {\bf NS fields}: The noncompact four
dimension index $\mu$ is not affected by the orbifold action, we
have ordinary NS fields,
$$\tilde\alpha_{-1}^\mu |0\rangle_\L b_{-1/2}^\nu |0\rangle_\R$$
whose traceless symmetric part, antisymmetric part and trace parts
are identified by gravition $g_{\mu\nu}$, antisymmetric tensor
$B_{\mu\nu}$ and dilaton $\phi$, respectively. There are also
superpartners, combined with the R sector of right movers.
$$\tilde\alpha_{-1}^\mu |0\rangle_\L d_0^{\nu} |0\rangle_\R$$
which has the same $\Z_3$ transformation property.  There are two
right moving Ramond states transforming as singlets, corresponding
two helicities. Since the fermionic statistics is inherited from
the right movers, the chiralities are also given by the R sector
of right movers. However we have Kaluza--Klein modes going around
the extra $i, \bar \imath$ direction, which transforms as triplet
and antitriplet under $\Z_3$.

\vskip 0.3cm\noindent{\bf Gauge fields}: Consider one $E_8$. The
massless modes of the first kind, satisfying $P\cdot V=$ integer
of (\ref{master}), include
\begin{center} \begin{tabular}{c}
$ \pm (0 ~ 1 ~ -1 ~ 0^5 ) $  \\
$ \pm (1 ~ \underline{1 ~ 0} ~ 0^5) $ \\
\end{tabular} \end{center}
and permutations of the underlined elements. With the two Cartan
generators, they form the root vectors ${\bf 8}$ of $SU(3)$. We
can identify the {\it simple roots} as $(-1~ -1 ~0~0^5)$ and
$(0~1~-1~0^5)$, since, from these, we can generate all the roots.
The inner products among the simple roots defines the Cartan
matrix $A$\index{Cartan matrix} and Dynkin diagram of $SU(3)$.
There exists another equivalent choice for positive roots and
simple roots.

We know that the remaining subgroup of $E_8$ is $E_6$.
\begin{center} \begin{tabular}{c}
$ (0~ 0~ 0~ \underline{ \pm 1 ~ \pm 1 ~ 0 ~ 0 ~ 0}) $ \\
$ \pm (- + + [+ + + + -]) $ \\
\end{tabular} \end{center}
and the square parenthesis denotes even sign flips. Here, the plus
and minus represents $+\frac12$ and $-\frac12$, respectively. We
can check that, together with the six $U(1)$ generators, they form
the root vectors ${\bf 78}$ of $E_6$. So, the first kind of the
untwisted sector determines the unbroken gauge group.

\vskip 0.3cm \noindent{\bf Matter fields}: Now we come to matter
representation. States satisfying $\exp (2\pi i P \cdot V) = \exp
2\pi i/3 \equiv \alpha$ are
\begin{center}\begin{tabular}{c}
$ (+\underline{+-}[++++-]) $ \\
$ (---[++++-]) $ \\
$ (0~\underline{1~0}~\underline{\pm 1 ~ 0^4}) $ \\
$ (-1~0~0~\underline{\pm 1 ~ 0^4}) $  \\
$ (1~\underline{-1~0}~0^5) $  \\
$ (0~-1~-1~0^5) $.  \\
\end{tabular}\end{center}
Under a $\Z_3$ rotation, these acquire a phase $\alpha$. We can
check that with an aid from the root vectors these form the weight
vectors $(\bf{3},\bf{27})$ of $SU(3)\times E_6$. The Hilbert space
$|L\rangle\otimes|R\rangle$ is completed with $|R\rangle$ of right
movers, transforming oppositely, so that they are invariant under
$\Z_3$. The remaining states, which transform like $\alpha^2$,
have the opposite quantum number and will be interpreted as
antiparticles. Thus, we require
\begin{equation}
P\cdot V=\frac13.\label{Umatter}
\end{equation}
This is good. We know that a ${\bf 27}$ contatins a complete
standard model family, including Higgses and right handed
neutrino.

\vskip 0.3cm \noindent{\bf $CTP$ conjugates}: The low energy field
inherit chirality from the right mover. So each matter field above
has only one four dimensional helicity $s_0$ when combined with
the R sector right movers such that it is $\Z_3$ invariant. Two
helicity states, $(\bf{3},{\bf 27})_-({\bf \overline 3},{\bf
\overline {27}})_+$, form a complete $(\bf{3},{\bf 27})$ in the
usual sense: $({\bf \overline 3},{\bf \overline {27}})_+$ being
the $CTP$ conjugate of $(\bf{3},{\bf 27})_-$. The multiplicity of
$(\bf{3},{\bf 27})$ is three because the right mover is a triplet
under $\Z_3$. The resulting spectrum is {\it chiral}.

\vskip 0.3cm \noindent{\bf Right mover}: For complete string state
we need right movers. The right movers are from an independent
superstring. The zero point energy is
 $0$ and $-\frac12$ for the R and the NS sectors, respectively.

In the NS sector, the ground state $b^M_{-1/2}|0\rangle$ has the
vector index $M=(\mu,i,\bar\imath)$. The transformation property
under the point group is the same as those of the left movers with
the same index. Those with spacetime index $\mu$ are invariant,
and those with holomorphic and antiholomorphic indices $i,\bar
\imath$ transform like $\alpha, \alpha^2$, respectively. This also
holds for bosonic degrees of freedom.

In the R sector, whose states are spinorial, the orbifold action
is not so clear that we devise the following. Bosonizing them, as
in the case for the description of the left moving group degrees
of freedom, we can represent them by a lattice vector $s
=(s_0~s_1~s_2~s_3)$ for $\bf{8}_{\rm s}$ and $\bf{8}_{\rm c}$.
Together with the vector representation, they are
\begin{eqnarray}
 \bf{8}_{\rm s} &=& ([++++]) \\
 \bf{8}_{\rm c} &=& ([-+++]) \\
 \bf{8}_{\rm v} &=& \pm(\underline{1~0~0~0})
\end{eqnarray}
which indeed coincide with the spinorial representation. This is
convenient because we can represent the NS and R states in a
unified way and we can deal the twisted sector states just as left
movers. By the GSO projection, we have the spinorial $\bf{8}_{\rm
s}$ only in the R sector and the vector $\bf{8}_{\rm v}$ in the NS
sector. Under the $\Z_3$ action $\phi$, each state is decomposed
as
\begin{eqnarray}
 \bf{3}: & (-+\underline{+-}),(----) &\sim\alpha^1 \\
 \overline{\bf{3}}: & (+-\underline{+-}),(++++) &\sim\alpha^2 \\
 \bf{1}: & (++--),(--++) & \sim\alpha^0
\end{eqnarray}
according to the transformation property of  $\phi \cdot s =
\alpha,\alpha^2,1,1$, respectively, for $\bf{3}+{\bf \bar 3+1+1}$.
In particular, the noncompact component $s_0$ remains untouched
and we adopt it as the helicity in four dimension.

In a $\Z_3$ invariant way, the singlet is combined with the left
moving gauge multiplet, and the triplet is combined with the
matter multiplet. The multiplicity three in the untwisted sector
matter is due to this triplet.

\vskip 0.3cm \noindent{\bf Twisted sector}:

\centerline{\it Left mover}

In the twisted sector, the zero point energy also shifts by
$f(\eta)$ for each real bosonic degree of freedom. In this case we
have four dimension with shift $\frac13$ and two with $\frac23$,
we have modified zero point energy to
$$ \textstyle \tilde c = 18 f(0) + 4 f(\frac13)
+ 2 f(\frac23)  = - \frac 23
$$
where $f(\eta)$ is given in Eq. (\ref{zeroptE}). Under the twist,
momentum shifts also by $V$. Thus the level matching condition
becomes
\begin{equation}\label{twistedex}
{M_\L^2 \over 4} = {(P+V)^2 \over 2} +\tilde\Nr - \frac23 =0.
\end{equation}
Without having the oscillator, $\tilde\Nr=0$ gives the condition
satisfying $(P+V)^2/2 = \frac23$, or $(P+V)^2=\frac43$, which is
satisfied by
\begin{center} \begin{tabular}{c}
$(0~-1~-1~0^5)$ \\
$(-1~0~0~\underline{\pm 1~0^4})$ \\
$(---[++++-]).$\\
\end{tabular} \end{center}
They transform as $(\bf{1},{\bf 27})$.

From the mode expansion of the twisted states, there are
fractional oscillators with $\tilde\Nr=\frac13$, from $\tilde
\alpha^i_{-1/3}$. So there are another set of states satisfying
$(P+V)^2/2=1/3$,
$$ \tilde \alpha^i_{-1/3}|P+V\rangle, \quad i=1,2,3, $$
with $P$
\begin{center} \begin{tabular}{c}
 $(0^8)$ \\
 $(-1~\underline{-1~0}~0^5)$ \\
\end{tabular} \end{center}
which form $(\bf \bar 3,1)$. Since we have $\chi(\phi)= 27$ fixed
point, we have three copies of states from the twisted sector due
to the oscillator $\tilde \alpha_{-1/3}$: $3 \times 27=81$ in
total.

\vskip 0.3cm \centerline{\it Right mover}

These states also combine with the right moving states. In the NS
sector, the zero point energy is
$c=2f(0)+2f(\frac23)+4f(\frac13)-[2f(\frac12)
+2f(\frac16)+4f(\frac56)]= 0$. The massless state is the ground
state $|0\rangle$.

The R sector has by definition vanishing zero point energy $c=0$.
Since the fields on the compact space(having index $i$ or $\bar
\imath$) are shifted by twisting, only $d^\mu_0|0\rangle$ are
massless. These states carry two helicities $s=\pm(----)$ one of
which is projected out by the GSO projection. These $s_i$ are
regarded as the bosonized coordinates. We just present the
condition \cite{FIQS},
\begin{equation}
 \Delta = \exp 2 \pi i [\tilde \Nr - \Nr
+(P+V)\cdot V-(s+\phi)\cdot \phi] =1,
 \label{rightprojector}
\end{equation}
The states having $(s+\phi) \cdot \phi=$ integer survive. The
twisted sector with shift $2\phi \simeq -\phi$ provides the
opposite helicity states.

In total we have
$$ 3 (\bf{3}, {\bf 27}) $$ in the
untwisted sector, and
$$ 27(\bf{1},{\bf 27}) + 81 ({\overline{\bf{3}}}, \bf{1})$$
in the twisted sector. In addition we have four dimensional ${\cal
N}=1$ supergravity multiplet coupled with SU(3)$\times$E$_6$ gauge
group and some Kaluza--Klein states. Since each ${\bf 27}$ has one
complete family, we have 36 chiral families.

It is a nontrivial check for the $SU(3)$
anomaly\index{anomaly!cancellation in $\Z_3$ orbifold}
cancellation between chiral fermions from the untwisted and
twisted sectors. $E_6$ itself is anomaly free, although it is a
complex representation.

\vskip 0.3cm\noindent {\bf Turning on background field: Another
shift vector from Wilson line}: When we have a compact dimension,
in general gauge field is not invariant globally, i.e. once we go
along the torus gauge field is invariant only up to a group
element. The piece that cannot be gauged away is
$$ A_I(x) = -i \Lambda^{-1} \partial_I \Lambda = \mbox{constant}. $$

\def\ket{\rangle}
\def\bra{\langle}
The gauge invariant measure of this constant gauge field is called
{\it Wilson line}\index{Wilson line},
$$ U = \exp \left[ 2\pi \oint A^I dx \cdot H^I \right]
    \equiv \exp \left(2 \pi a^I \cdot H^I \right). $$
Here, we made group space index $(I=1,\dots,16)$ explicit.
Equivalently,
\begin{equation}
 | P \ket \to U | P \ket = \exp (2 \pi i a \cdot P) | P \ket.
\end{equation}
This is nothing but a shift vector. In general the Wilson line can
assume continuous values, however we choose discrete values in
order to associate with the orbifold action. Up to now we have
considered only the gauge embedding of the kind $(\theta,0) \to
V$. We also have another possibility of the lattice translation
$(1,v) \to a$, since this is another action on orbifold. The
Wilson line shift vector must be consistent with the orbifolding:
$N a$ is also belonging to the weight lattice\index{order} and
should satisfy the modular invariance condition.

Now we take into account such translations.  Recall that the point
group action
\begin{equation}
 X^i(\sigma+\pi) = (\theta X)^i(\sigma) + \sum_{a=1}^6 m_a
e^i_a \label{record}
\end{equation}
with $e^i_a$ being unit vector along the $a^{\rm th}$ torus. So
far we have neglectd the translational piece because all the fixed
point have been equivalent: we have no way to tell the specific
fixed point. However when we turn on the background field, we see
a effect whenever we go along the torus. Coming back to the
original point by (\ref{record}), we have the effect of Wilson
line
$$
X^I(\sigma+\pi) = X^I (\sigma) + V^I + \sum_{a=1}^6 m_a a_a^I
$$
so that we have a homomorphism
$$ \theta \to V, \quad e_a \to a_a, $$
to have. This is equivalent to the shift in the momentum space
$$ P^I \to P^I + V^I + \sum_{a=1}^6 m_a a_a^I. $$
In the untwisted sector, the projection is modified by additional
piece $a_a$. The net effect in the untwisted sector is the common
intersection, leading
\begin{equation}
P \cdot a_a = \mbox{integer}, \quad \mbox{for all $a$}.
\end{equation}
In the twisted sector, the mass shell condition becomes
$$ {M_\L^2 \over 4} = {(P^I + V^I + \sum_a m_a a_a^I)^2 \over 2}
+ \tilde \Nr + \tilde c,
$$
noting that $m_a$ determines the specific fixed point. The $\tilde
c$ is the normal ordering constant from the internal field
oscillators; thus it remains the same as that without the Wilson
line. With this in mind, we modify the modular invariance
condition  to
\begin{equation}
 (N \phi)^2 = \left[ N (V+\sum m_a a_a) \right]^2 = 0 \label{modinvwilson}
\end{equation}
where the vector indices are suppressed. In the simplest $\Z_3$
case, it reduces to
\begin{equation}
V \cdot a_a ={\rm integer},\ \ \ a_a \cdot a_b ={\rm integer,\ \ \
for\ all\ } a_a
\end{equation}

There can exist nontrivial relations between shift vectors. For
instance, in the SU(3) lattice of the preceding section, a unit
lattice vector transforms like $\theta e_a = e_{a+1}$. Thus, the
Wilson line along this lattice should be the same
\begin{equation}
a_{a+1} = a_{a}.
\end{equation}
Namely, there can exist only three independent Wilson lines when
we compactify 6 dimensions. If Wilson lines $a_a$ and $a_{a+1}$
are not related by the lattice transformation, they need not be
related. For instance, for a $\Z_4$ orbifold $\phi=\frac14(2~1~1)$
the action on the first two-torus is the reflection about the
origin, i.e. $\Z_2$. In this case there is no relation between two
unit vectors defining the first two-torus, hence the Wilson line
shift vectors $a_1$ and $a_2$ are independent. It means that we
can take the two-torus with the edges having different sizes, $R_1
\ne R_2$.

\vskip 0.3cm\noindent {\bf $\Z_3$ example with Wilson lines}:
There are three fixed points in each two-torus. Under the point
group action, they return to original point, only by accompanied a
translation $(0,0),(1,0),(0,1)$, respectibely. The `square'
remains invariant, but the `cross' need $e_4$ translation after
the point action. By the $\Z_3$ symmetry $\theta e^4 = e^5$, we
have automatically $a_2=a_1$ along the $X^5$ direction. So we do
not need to check the condition from $a_2$.

Now we add a Wilson line $a_1$ along the $X^4$ direction,
\begin{equation}\label{a1example}
\textstyle a_1 = \frac13(5~1^5~0~0)(0^8).
\end{equation}
One can easily check that this satisfies the modular invariance
condition (\ref{modinvwilson}).

The untwisted sector fields are common intersection of those
obtained by resorting to the conditions $P\cdot V=$ integer and
$P\cdot a_1=$ integer. The resulting untwisted spectrum is those
arising from gauge group $SU(3)^4$:
$(8,1,1,1)+(1,8,1,1)+(1,1,8,1)+(1,1,1,8)$ and three chiral
$(3,\bar 3,1,3)$. The multiplicity 3 is due to the triplet nature
of the right moving states.

The center of mass of twisted string is at the fixed point.
$$ X^i(2 \pi) = \theta X^i (0) + \sum_{a=1}^6 m_a e_a^i .$$
With the Wilson line, the mass shell condition of the twisted
sector spectrum is,
\begin{equation}\label{TSWilson}
{M_\L^2 \over 4} = {(P+V+\sum_{a} m_a a_a^1)^2 \over 2} + \tilde
\Nr + \tilde c
\end{equation}
where the summation on $a$ is $a=1,2,\cdots,6$. The reference
\cite{Kim03} contains all the roots and weight lattices in terms
of shift vectors.

It is obvious that the Wilson lines are {\it non-contractible
loops} because they belong to shifts.
\begin{figure}[t]
\begin{center}
\begin{picture}(400,90)(0,-20)

\SetWidth{0.5} \Oval(74,24)(40,50)(0)\Curve{(45,35)(70,20)(95,35)}
\Curve{(50,30)(72,38)(90,30)}

\SetWidth{1.2} \Line(103.33,0)(99.34,-4) \Line(99.34,0)(103.34,-4)
\Text(90,20)[c]{$\bullet$} \GBox(57.4,-1.6)(61.6,1.6){0}

\SetWidth{0.5}
\Oval(189,24)(40,50)(0)\Curve{(160,35)(185,20)(210,35)}
\Curve{(165,30)(187,38)(205,30)}

\SetWidth{1.2} \Line(218.33,0)(214.34,-4)
\Line(214.34,0)(218.34,-4) \Text(205,20)[c]{$\bullet$}
\GBox(171.4,-1.6)(176.6,1.6){0}

\SetWidth{0.5}
\Oval(304,24)(40,50)(0)\Curve{(275,35)(300,20)(325,35)}
\Curve{(280,30)(302,38)(320,30)}

\SetWidth{1.2} \Line(333.33,0)(329.34,-4)
\Line(329.34,0)(333.34,-4) \Text(320,20)[c]{$\bullet$}
\GBox(287.4,-1.6)(291.6,1.6){0}

\end{picture}
\caption{The torus and the fixed points of the ${\bf Z}_3$
orbifold. }\label{z3fixed}
\end{center}
\end{figure}
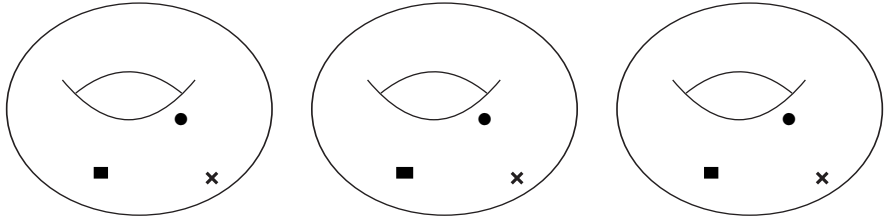
It should be noted that, in the presence of Wilson line, 27 fixed
points are not equivalent any more. This is because, by the
lattice shift $m_a$ connecting endpoints of the twisted string,
strings located at different fixed points feel Wilson lines by
different amounts, no wrapping, + direction, or -- direction, as
discussed above, and they are shown in Fig. \ref{z3fixed} with
$a_1=a_2$ and $a_3=a_4$. Each separated fixed point carries a
different Wilson line in general.

In Fig. \ref{z3fixed}, $a_1$ in the first torus distinguishes the
three fixed points in the first torus and $a_3$ in the second
torus distinguishes the three fixed points in the second
torus.\footnote{The Wilson line shown around a fixed point is in
fact two winding Wilson lines in the opposite directions, cut and
patched together. In the calculation, we use the winding Wilson
lines.}

The product of the first and the second torus gives 9 distinct
fixed points. Thus, for the chiral fermions in the $Z_3$ orbifold,
we compute the massless spectrum from these 9 different twisted
sectors in addition to the untwisted sector $U$. But the fixed
points of the third torus is not distinguished,
 and hence each of the nine twisted sectors has the
multiplicity 3. Thus, the Wilson line dictates which field lives
at which fixed point. In field theoretic models, there is no
corresponding restriction of this behavior of the string twisted
sector, and hence there is no principle in field theory for the
fields at the fixed points except for the less strict requirement
of the anomaly cancelation.

\section{Need for HESSNA: Trinification example}

The $\Z_3$ orbifolds has multiplicity 3 for the U fields and
multiplicities at least 3 for the T fields. With two Wilson lines,
every sector has multiplicity 3. Thus, we concentrate on two
Wilson line models for the magic family number 3 from the outset.
In this spirit, $\Z_3$ orbifolds toward standard-like models was
initiated long time ago \cite{iknq}, which achieved obtaining, (i)
the correct SM gauge group in SU(3)$\times$SU(2)$\times$U(1)$^n$,
(ii) 3 families, and (iii) no extra color triplets in some models
the so-called doublet-triplet splitting. But, in this kind of
standard-like models there exist two serious problems: (1) the
\sw0\ problem and (2) the problem of too many Higgs doublets. The
problem (1) is to be resolved if we want to keep the coupling
constant unification and it was argued that semi-simple groups
with rank $\ge 5$ are needed \cite{kimsw0}. We do not attempt to
answer (2) at this moment.

The basic reason for the \sw0\ problem is that the electroweak
hypercharge $Y$ is leaked to the uncontrollably many U(1)s. GUTs
are good but the difficulty with GUTs is that the adjoint
representations do not appear at the Kac-Moody level 1. This is
the reason for the hypercharge embedding in semi-simple groups
with no need for an adjoint representation(HESSNA). The possible
candidates are the Pati-Salam type group and the trinification
group. For the simplest orbifold $\Z_3$, the trinification is the
natural possibility.

\subsection{Group theory on $Z_N$ embedding}

To classify possible groups from $\Z_N$ orbifolds, it was pointed
out that the Dynkin diagram techniques is quite useful \cite{chk}.
Using this techniques, the program has been developed
\cite{Hwang}, and all the $\Z_3$ orbifolds with two Wilson lines
are tabulated \cite{BookOrb}.

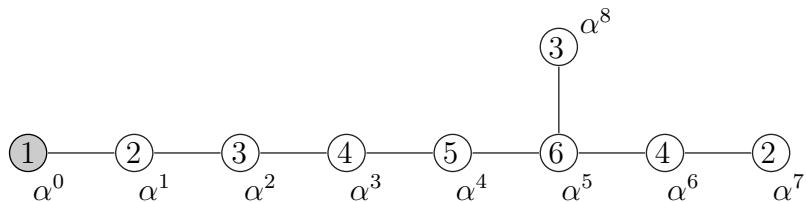
\begin{figure}[h]
\begin{center}
\begin{picture}(400,80)(0,-20)

\SetWidth{0.5}
 \GOval(40,0)(7,7)(0){0.8}
\Text(40,0)[c]{1} \Text(80,0)[c]{2}\Text(120,0)[c]{3}
\Text(160,0)[c]{4}\Text(200,0)[c]{5}\Text(240,0)[c]{6}
\Text(280,0)[c]{4}\Text(320,0)[c]{2}\CArc(80,0)(7,0,360)\CArc(120,0)(7,0,360)
\CArc(160,0)(7,0,360)\CArc(200,0)(7,0,360)\CArc(240,0)(7,0,360)
\CArc(280,0)(7,0,360)\CArc(320,0)(7,0,360)
\Text(48,-13)[c]{$\alpha^0$}\Text(88,-13)[c]{$\alpha^1$}
\Text(128,-13)[c]{$\alpha^2$}
\Text(168,-13)[c]{$\alpha^3$}\Text(208,-13)[c]{$\alpha^4$}
\Text(248,-13)[c]{$\alpha^5$}
\Text(288,-13)[c]{$\alpha^6$}\Text(328,-13)[c]{$\alpha^7$}

\Line(47.5,0)(72.5,0)\Line(87.5,0)(112.5,0)
\Line(127.5,0)(152.5,0)
\Line(167.5,0)(192.5,0)\Line(207.5,0)(232.5,0)
\Line(247.5,0)(272.5,0) \Line(287.5,0)(312.5,0)

\Text(240,40)[c]{3}\CArc(240,40)(7,0,360)
\Text(255,50)[c]{$\alpha^8$} \Line(240,7.5)(240,32.5)

\end{picture}
\caption{Extended Dynkin diagram of E$_8$ and Coxeter labels shown
inside circles. The extended root is grey colored.}\label{e8dyn}
\end{center}
\end{figure}

The largest group we consider is E$_8$ from the \E88\ heterotic
string. Its extended Dynkin diagram with the Coxeter levels are
shown in Fig. \ref{e8dyn}.
The Coxeter level is used for obtaining gauge groups of $\Z_N$
orbifolds which is studied with Wilson lines in Ref. \cite{chk}.

\subsection{Trinification from superstring}

With the orbifold techniques discussed in this lecture, we can
question, ``Is there any interesting trinification model?" As one
can see from the tables of \cite{BookOrb}, there are many
trinification models. But supersymmetric trinification models must
hurdle over the R-parity problem and the neutrino mass problem
\cite{cchk}.  Most trinification models do not solve these, but
there is one interesting model solving these two problems using
some linkage fields \cite{kimtri}. Without  such linkage fields,
there is no acceptable trinification model from string orbifolds.

\paragraph{Acknowledgments} JEK thanks all organizers of SI-2004, and
especially Prof. Morimitsu Tanimoto, for inviting him for this talk
and to  beautiful Mt. Fuji. This work is supported in part by the
KOSEF Sundo Grant, the ABRL Grant No. R14-2003-012-01001-0, and the
BK21 program of Ministry of Education, Korea.


\end{document}